\documentclass[a4paper,11pt]{article} 
\usepackage{authblk}
\usepackage{fullpage}
\usepackage{graphicx}
\usepackage{xcolor}
\usepackage{algorithm}
\usepackage{algpseudocode}
\usepackage{amssymb}
\usepackage{amsthm}
\usepackage{amsmath}
\usepackage{mathrsfs}
\usepackage[mathcal]{euscript}
\usepackage{bm}
\usepackage{mathabx}
\usepackage[colorlinks,urlcolor=cobalt,citecolor=cobalt,linkcolor=cobalt,pdftex, pdfauthor = Lukas Exl]{hyperref}
\usepackage{tabularx}
\usepackage{setspace} 
\usepackage{wasysym}

\usepackage{ulem}

\definecolor{airforce}{rgb}{0.16,0.32,0.75}
\definecolor{cobalt}{rgb}{0.0,0.28,0.67}

\newcommand*\luk[2]{{\color{black}{#2}}} 

\newtheorem{thm}{Theorem}
\newtheorem{lem}[thm]{Lemma}

\newtheorem{cor}[thm]{Corollary}
\newtheorem{defi}{Definition}

\title{\Large{\textbf{Prediction of magnetization dynamics in a reduced dimensional feature space setting utilizing a low-rank kernel method}}} 
\author[1,4]{Lukas Exl \thanks{\texttt{lukas.exl@univie.ac.at}}}
\author[1,4]{Norbert~J.~Mauser}
\author[1,4]{Sebastian Schaffer}
\author[2,4]{Thomas Schrefl}
\author[3,4]{Dieter Suess}
\affil[1]{\small Wolfgang Pauli Institute c/o Faculty of Mathematics, University of Vienna, Austria.} 
\affil[2]{Christian Doppler Laboratory for Magnet design through physics informed machine learning, Department of Integrated Sensor Systems, Danube University Krems, Austria}
\affil[3]{Faculty of Physics, University of Vienna, Austria}
\affil[4]{University of Vienna Research Platform MMM Mathematics - Magnetism - Materials, University of Vienna, Austria}

\begin{document}
%
\maketitle
\date

\noindent\textbf{Abstract.} We establish a machine learning model for the prediction of the magnetization dynamics as function of the external field described by the Landau-Lifschitz-Gilbert equation, 
the partial differential equation of motion in micromagnetism. The model allows for fast and accurate determination of the response to an external field which is illustrated by a thin-film standard problem. 
The data-driven method internally reduces the dimensionality of the problem by means of nonlinear model reduction for unsupervised learning. This not only makes accurate prediction of the time steps possible, but also 
decisively reduces complexity in the learning process where magnetization states from simulated micromagnetic dynamics associated with different external fields are used as input data. 
We use a truncated representation of kernel principal components to describe the states between time predictions. 
The method is capable of handling large training sample sets owing to a low-rank approximation of the kernel matrix and an associated low-rank extension of kernel principal component analysis and kernel ridge regression. The approach entirely shifts computations  
into a reduced dimensional setting breaking down the problem dimension from the thousands to the tens.\\  

\noindent\textbf{Keywords.} nonlinear model order reduction, low-rank kernel principal component analysis, Nystroem approximation, low-rank kernel approximation, machine learning, micromagnetics\\

\noindent\textbf{Mathematics Subject Classification.} 	62P35,\, 68T05,\,  65Z05

\newpage
\section{Introduction}
Computational micromagnetics is a broad scientific field with useful technological applications such as permanent magnets \cite{fischbacher2018micromagnetics} or magnetic sensors \cite{suess2018topologically}. The dynamics of the magnetization in a magnetic material influenced by internal and external fields is mathematically described by the Landau-Lifschitz-Gilbert (LLG) equation, a time-dependent partial differential equation (PDE). The numerical challenge involves many time-consuming computations of solutions to a Poisson equation in whole space \cite{abert2013numerical,exl2018magnetostatic} for evaluating derivatives in the course of the time-stepping scheme \cite{miltat2007numerical,schrefl2007numerical}. 
In contrast, electronic circuit design and real time process control need models that provide the sensor response quickly. 
A way to provide such demands for applications is offered by diverse reduced order models (ROMs) in micromagnetism. So far most ROMs were (multi)linear, e.g., tensor methods \cite{exl2014thesis} and model reduction based on spectral decomposition \cite{bruckner2019large} such as via a subset of the eigenbasis of the discretized self-adjoint effective field operator \cite{d2009spectral}. While these are keen ideas, they are clearly limited due to the inherent linearity of the reduced models. Recently, the authors introduced data-driven nonlinear model order reduction (nl-MOR) to effectively predict the magnetization LLG-dynamics subject to the external field based on simulated data \cite{kovacs2019learning,exl2020learning}. 
Fast response to an external field can be obtained from such data-driven PDE machine learning (ML) models combined with unsupervised nonlinear model reduction. Another inspiration of the proposed machine learning scheme in \cite{exl2020learning} was to construct a time-stepping predictor on the basis of a non-black-box nonlinear dimensionality reduction approach such as kernel principal component analysis (kPCA) \cite{scholkopf1997kernel} for the better understanding of the underlying approximations. In this context, the key idea is to use a data set of simulated magnetization trajectories to learn a time-stepping model scheme that is capable of predicting the dynamics step by step for a new unseen external field without having to solve the LLG equation numerically, and hence with practically negligible computational effort. The challenging part is the combination of the learning process with reduced dimensionality of the feature space, which is initially proportional to the size of the discretization space used in the data generation, thus, several orders of magnitude too large for regression. A nonlinear kernel version of principal component analysis for the feature space dimensionality reduction was successfully established in \cite{exl2020learning}, where each time-step was learned on the basis of magnetization states represented via truncated kernel principal components. In the forthcoming presentation one novel extension of the idea in \cite{exl2020learning} will be the simultaneous learning of all steps via an entire dimension-reduced feature space integration scheme. Besides, the second improvement concerns the feasibility of the kernel learning scheme by the introduction of low-rank approximation to the kernel matrix, which allows the use of larger learning data. The reason for its importance is the fact that the learning process gets gradually infeasible as data size increases, as such, a common problem in data-driven methods but especially the case for kernel methods in machine learning \cite{hofmann2008kernel}. Thus, while the original approach already leads to an exceptional reduction in feature dimension and fast learning thanks to the nonlinear kernel, the novel approach performs training and predictions entirely in reduced coordinates and is capable of exploiting information from large training data sample sets owing to the low-rank kernel principal component analysis (low-rank kPCA).  

The paper is structured as follows. First we give a brief overview of the ML approach for learning maps between feature spaces 
with reduced dimensionality. The section starts with an introduction to kernels and kernelized principal component analysis. Following this, we introduce the low-rank approximations of the kernel methods including kPCA, kernel ridge regression (kRR) and the crucial pre-image computation. The low-rank method is validated by means of an example from the scikit library \cite{scikitlearn}. In the end of the method section~\ref{sec:methods} we give the general procedure for learning maps between feature space elements with truncated components. Section~\ref{sec:numerics} covers the application to micromagnetics including decription of data (structure) as well as several numerical validations based on a standard problem \cite{mcmichael_mag_nodate}.       

\section{Learning feature space maps with reduced dimensionality}\label{sec:methods}
%
In the following we give a brief description of \textit{kernels} and \textit{feature spaces as reproducing kernel Hilbert spaces (RKHS)}. 
A more comprehensive discussion on the core definitions of general kernel methods can be found for instance in the review \cite{hofmann2008kernel}. 
Central to our approach is \textit{kernel principal component analysis (kPCA)} as a means for unsupervised learning and model reduction. We will extend kPCA to its \textit{low-rank variant} 
to be able to handle large data sets effectively in learning \textit{feature space maps}. Moreover, we establish a \textit{low-rank kernel ridge regression} (low-rank kRR) for large training data and effective 
pre-image computation.  
\subsection{Kernel principal component analysis}
The definition of a kernel function is given as follows.
\begin{defi}[Positive definite kernel function]\label{def:kernel} 
 Let $\mathcal{X}$ be a nonempty set. A symmetric function $k:\, \mathcal{X} \times \mathcal{X} \rightarrow \mathbb{R}$ is a positive definite kernel on $\mathcal{X}$ if 
 for all $m\in \mathbb{N}$ any choice of inputs $\mathbf{x} = \{x_1,\hdots,x_m\} \subseteq \mathcal{X}$ gives rise to a positive definite gram matrix $K[\mathbf{x}] \in \mathbb{R}^{m \times m}$ 
 defined as $K_{ij} = k(x_i,x_j),\,i,j = 1,\hdots,m$. To distinguish an involved second subset $\mathbf{y} = \{y_1,\hdots,y_\ell\}  \subseteq \mathcal{X}$ we define the matrix $K[\mathbf{x},\mathbf{y}] \in \mathbb{R}^{m \times \ell}$ via its entries 
 $K_{ij} = k(x_i,y_j),\,i = 1,\hdots,m,\, j=1,\hdots,\ell$. 
\end{defi}
\noindent We will refer to positive definite kernels as \textit{kernels}. An important class of kernels are the \textit{Gaussian kernels} also known as \textit{(Gaussian) radial basis functions (RBF)}.
\begin{defi}[RBF]\label{def:rbf}
 Let $\mathcal{X}$ be a dot product space. The radial basis function (RBF) kernel between two vectors $x,y \in \mathcal{X}$ is defined as 
 \begin{align}
  k(x,y) = e^{-\gamma \|x-y\|^2}.
 \end{align}
For the choice $\gamma = 1/\sigma^2$ the kernel $k$ is also known as the Gaussian kernel of variance $\sigma^2$.
\end{defi}

Kernels represent a way to express \textit{similarity measures} and can be used to extend linear structural analysis for data like the (linear) PCA to nonlinear analogues. 
Mathematically, a possibly infinite dimensional \textit{Hilbert space} $\mathcal{F}_k$ can be constructed, called the \textit{feature space} of $\mathcal{X}$ associated with the kernel $k$, 
where the inner product is defined by the kernel $k$. A (nonlinear) map $\phi_k:\,\mathcal{X} \rightarrow \mathcal{F}_k$ ''embeds'' the data in the feature space, i.e., $\phi_k(\mathcal{X}) = \mathcal{F}_k$. 
$\mathcal{F}_k$ is a RKHS and mathematically well understood, e.g., see \cite{saitoh1988theory} for the theoretical background. 
It is important to note that the inner product in $\mathcal{F}_k$ of two mapped data points can be computed without knowledge of the map $\phi_k$ as 
\begin{align}\label{eqn:kerneltrick}
 \phi_k(x) \cdot \phi_k(y) = k(x,y),
\end{align}  
which is known as the \textit{kernel trick} in the machine learning community.
Intuitively, a mapped data point can be seen as a new vector $\phi_k(x) = (p_1(x),p_2(x),\hdots)^T$ where the nonlinear functions $p_j$ define 
the coordinates of $\phi_k(x)$ in the higher dimensional feature space. One can now try to learn structure via the mapped inputs by extending linear algorithms for unsupervised learning, like the PCA, to operate on the 
feature space. This is known as \textit{kernelization} where the kernelized version of the linear PCA is known as the \textit{kernel principal component analysis} (kPCA) \cite{scholkopf1997kernel}.
The algorithm of kPCA is given next.
\begin{defi}[kPCA]\label{def:kpca}
 Given inputs $\mathbf{x} = \{x_1,\hdots,x_m\} \subseteq \mathcal{X}$ and a kernel $k:\, \mathcal{X} \times \mathcal{X} \rightarrow \mathbb{R}$ the kernel PCA generates 
 kernel principal axes $v^{(j)} = \tfrac{1}{\sqrt{\lambda_j}} \,\sum_{i=1}^m \alpha_i^{(j)} \phi_k(x_i),\, j=1,2,\hdots,m,$ where the coefficient vectors $\alpha^{(j)} \in \mathbb{R}^{m},\,j=1,\hdots,m$ result from the eigenvalue problem 
 \begin{align}\label{eqn:eigkpca}
  G\alpha^{(j)} = \lambda_j \alpha^{(j)},
 \end{align}
where the centered gram matrix $G = K[\mathbf{x}] - \mathbf{1}_m K[\mathbf{x}] - K[\mathbf{x}]\mathbf{1}_m + \mathbf{1}_mK[\mathbf{x}]\mathbf{1}_m \in \mathbb{R}^{m \times m}$ with $(\mathbf{1}_m)_{ij} = 1/m$ is used. 
The eigenvalue problem \eqref{eqn:eigkpca} is solved for nonzero eigenvalues. 
The $j$-th kernel principal component of a data point $x \in \mathcal{X}$ can be extracted by the projection  
\begin{align}
 p_j(x) = \phi_k(x) \cdot v^{(j)} = \frac{1}{\sqrt{\lambda_j}} \sum_{i=1}^m \alpha_i^{(j)} k(x_i,x).
\end{align}
\end{defi}
For the purpose of \textit{nonlinear dimensionality reduction} only a few kernel principal components $p_j(x)$ are extracted.\\
The problem of finding pre-images of kPCA components is mathematically challenging. If Gaussian kernels are used, this can be done by the use of fixed point iterations, while a general purpose method, 
which proves to be practically reliable, is to learn the pre-images during the process of establishing the kPCA model through the training data \cite{bakir2004learning}. We will describe our approach to pre-image computation within our low-rank framework in the forthcoming section. 
%
%
\subsection{Low-rank kernel principal component analysis}\label{sec:lowrankkpca} 
For large sample size $m$ we seek a low-rank approximation of the Gram matrix as a Nystroem approximation of the kernel matrix arising from the training data split into $r \leq m$ randomly selected basis samples and the $m-r$ remaining samples \cite{williams2001using}.\\ 
In the case where the kernel matrix has rank $r\leq m$ we get an explicit form of the low-rank decomposition.
\begin{lem}[Low-rank approximation of the kernel matrix]\label{lem:lowrank}
 Given samples $\mathbf{x} = \{x_1,\hdots,x_m\} \subseteq \mathcal{X}$ we assume to be able to pick a subset of $r \leq m$ samples $\mathbf{x}_r \subseteq \mathbf{x}$ such that $K[\mathbf{x_r}] \in \mathbb{R}^{r \times r}$ 
 has full rank $r$. Let us further denote the set of remaining $m-r$ samples with $\mathbf{x}_{m-r}$ and assume the relabeled initial sample set $\mathbf{x} = \{\mathbf{x}_r, \mathbf{x}_{m-r}\}$ such that the associated kernel matrix 
 gets block form
 \begin{align}
  K[\mathbf{x}] = \left(
   \begin{array}{c c}
    K_{r,r} & K_{m-r,r}^T \\
    K_{m-r,r} & K_{m-r,m-r}
   \end{array}\right),
 \end{align}
where $K_{r,r} := K[\mathbf{x}_r] \in \mathbb{R}^{r \times r}$, $K_{m-r,r} := K[\mathbf{x}_{m-r},\mathbf{x}_{r}] \in \luk{}{\mathbb{R}^{(m-r) \times r}}$ and $K_{m-r,m-r} := K[\mathbf{x}_{m-r}] \in \mathbb{R}^{(m-r) \times (m-r)}$.
Then there holds 
\begin{align}
 K[\mathbf{x}] = \Phi_r\,\Phi_r^T,
\end{align}
with 
\begin{align}\label{eqn:phi}
 \Phi_r := \Phi_r[\mathbf{x}] = \left(
 \begin{array}{c}
  K_{r,r}^{1/2}\\
  K_{m-r,r}\,K_{r,r}^{-1/2}
 \end{array}\right) = 
 K_{m,r}\,K_{r,r}^{-1/2} \in \mathbb{R}^{m \times r}.
\end{align}
\end{lem}
\noindent\textit{Proof.}\quad We first observe that 
\begin{align}
\Phi_r\,\Phi_r^T = \left(
   \begin{array}{c c}
    K_{r,r} & K_{m-r,r}^T \\
    K_{m-r,r} & K_{m-r,r}K_{r,r}^{-1}K_{m-r,r}^T
   \end{array}\right).
\end{align}
Since $K[\mathbf{x}]$ has rank $r$, we have the eigenvalue decomposition $K[\mathbf{x}] = U \Lambda U^T$ with $U \in \mathbb{R}^{m\times r}$ and the diagonal matrix 
$\Lambda \in \mathbb{R}^{r \times r}$ built from the $r$ nonzero eigenvalue of $K[\mathbf{x}]$. Using the block notation $U = (U_{r}^T,U_{m-r}^T)^T$ we get 
\begin{align}
 K[\mathbf{x}] =  U \Lambda U^T = \left(
 \begin{array}{c c}
    U_{r}\Lambda U_{r}^T & U_{r}\Lambda U_{m-r}^T \\
    U_{m-r}\Lambda U_{r}^T & U_{m-r}\Lambda U_{m-r}^T
   \end{array}\right).
\end{align}
Note that $U_r^T U_r = I$ and hence
\begin{align}
 K_{m-r,r}K_{r,r}^{-1}K_{m-r,r}^T =  (U_{m-r}\Lambda U_{r}^T)   (U_{r}\Lambda^{-1} U_{r}^T)  (U_{r}\Lambda U_{m-r}^T) = U_{m-r}\Lambda U_{m-r}^T = K_{m-r,m-r},
\end{align}
which shows $K[\mathbf{x}] = \Phi_r\,\Phi_r^T$. Finally, the identity in  Eqn.~\eqref{eqn:phi} simply follows from $K_{r,r}^{1/2} = K_{r,r}K_{r,r}^{-1/2}.$ \hfill $\Square$\\

Note that the computation of $\Phi_r$ only needs $\mathcal{O}(mr + r^2)$ kernel evaluations and additional cost of $\mathcal{O}(r^3)$ for the root $K_{r,r}^{-1/2}$ plus a cost of $\mathcal{O}(mr^2)$ for the matrix multiplication.\\
We further remark that for some kernels such as Gaussian RBF the above rank $r$ assumption will only hold approximately for sufficiently large $r$.\\  

From Lemma~\ref{lem:lowrank} Eqn.~\eqref{eqn:phi} we see that when mapping an individual data sample $y \in \mathcal{X}$ under $\Phi_r:\,\mathcal{X}\rightarrow \mathbb{R}^r$ the corresponding feature vector is given as 
\begin{align}
 \Phi_r(y) = \big( k(y,x_1),\hdots,k(y,x_r) \big)\,K_{r,r}^{-1/2},
\end{align}
which holds true for $y \notin \mathbf{x}_r$ as it represents the respective row in $K_{m,r}\,K_{r,r}^{-1/2}$, but also for $y \in \mathbf{x}_r$ due to the identity $K_{r,r}^{1/2} = K_{r,r}K_{r,r}^{-1/2}$. 
We summarize this remark for later reference. 
\begin{cor}\label{cor:phiy}
 Under the assumptions of Lemma~\ref{lem:lowrank} the matrix of feature vectors of data samples $\mathbf{y} = \{y_1,\hdots,y_\ell\}$ is given as
 \begin{align}
  \Phi_r[\mathbf{y}] = K[\mathbf{y},\mathbf{x}_r] \, K_{r,r}^{-1/2},
 \end{align}
with $ K[\mathbf{y},\mathbf{x}_r] = \big( k(y_i,x_j)_{i,j} \big)\in \mathbb{R}^{\ell \times r}$. 
\end{cor}

In the course of the kPCA algorithm one has to solve $d$ eigenvalue problems of the form $G\alpha^{(j)} = m \lambda_j \alpha^{(j)},\,j = 1,\hdots,d$ which now take the particular form 
\begin{align}\label{eqn:lowrankeig}
 \bar{\Phi}_r\bar{\Phi}_r^T \alpha^{(j)} =  \lambda_j \alpha^{(j)},\,j = 1,\hdots,d,
\end{align}
with $\bar{\Phi}_r = \Phi_r- \mathbf{1}_m \Phi_r$ and $\Phi_r = \Phi_r[\mathbf{x}] = K_{m,r} K_{r,r}^{-1/2}\in \mathbb{R}^{m \times r}$ with $K[\mathbf{x}] \approx \Phi_r[\mathbf{x}]\Phi_r[\mathbf{x}]^T$ being the low-rank approximation 
of the kernel matrix.  
An eigenvalue problem of the form \eqref{eqn:lowrankeig} can be efficiently solved for nonzero eigenvalues.  
\begin{lem}[Low-rank eigenvalue problem]\label{lowrankeig} The eigenpairs $(v,\lambda )$ with $\lambda \neq 0$ of $\Phi_r \Phi_r^T \in \mathbb{C}^{m \times m}$ with $\Phi_r \in\mathbb{C}^{m \times r}$ are given by $(\Phi_r w, \lambda)$ with $\Phi_r^T \Phi_r w = \lambda w$. Particularly, there holds for $\|w\|_2 = 1$ that $\|v\|_2 = \lambda^{1/2}$, i.e., $v = \lambda^{-1/2}\,\Phi_r w$ has unit length.
\end{lem}
\noindent\textit{Proof.}\quad Suppose $\Phi_r^T \Phi_r w = \lambda w$ with $\lambda \neq 0$. Then we have $\Phi_r \Phi_r^T (\Phi_r w) = \lambda (\Phi_r w)$ with  $\Phi_r w \neq 0$, since otherwise multiplication with $\Phi_r^T$ yields $\Phi_r^T \Phi_r w = 0$ and thus, $\lambda = 0$, contradicting the assumption $\lambda \neq 0$ in the first place. Hence, $(\Phi_r w,\lambda)$ is an eigenpair of $\Phi_r \Phi_r^T$. Moreover, $\|\Phi_r w\|_2^2 = w^T (\Phi_r^T \Phi_r w) = \lambda w^Tw = \lambda.$ \hfill $\Square$\\[0.1cm]
The remarkable consequence of Lemma~\ref{lowrankeig} is a significant reduction in complexity when solving the eigenvalue problems in the kPCA with low-rank kernel matrix approximation in the case $r \ll m$. 
Specifically, the computational complexity is reduced from \luk{}{$\mathcal{O}(m^2)$ to $\mathcal{O}(r^2)$} for each of the $d$ eigenvalue problems.
We now have the tools to define a low-rank version of the kPCA.

\begin{defi}[Low-rank kPCA]\label{def:lowrankkpca}
 Given inputs $\mathbf{x} = \{x_1,\hdots,x_m\} \subseteq \mathcal{X}$, a kernel $k:\, \mathcal{X} \times \mathcal{X} \rightarrow \mathbb{R}$ and a low-rank 
 approximation of $G = K[\mathbf{x}] - \mathbf{1}_m K[\mathbf{x}] - K[\mathbf{x}]\mathbf{1}_m + \mathbf{1}_mK[\mathbf{x}]\mathbf{1}_m \in \mathbb{R}^{m \times m}$ by 
 $\bar{\Phi}_r\bar{\Phi}_r^T = (\Phi_r- \mathbf{1}_m \Phi_r)(\Phi_r- \mathbf{1}_m \Phi_r)^T$ from a choice of a subset $\mathbf{x}_r \subseteq \mathbf{x}$ according to Lemma~\ref{lem:lowrank}. The low-rank version of the kernel PCA generates 
 $r \leq m$ kernel principal axes $v^{(j)} = \tfrac{1}{\sqrt{\lambda_j}} \,\sum_{i=1}^m \alpha_i^{(j)} \bar{\Phi}_r(x_i),\, j=1,2,\hdots,r,$ where the coefficient vectors $\alpha^{(j)} \in \mathbb{R}^{m}$ result from the eigenvalue problem 
 \begin{align}\label{eqn:eigkpca2}
  \bar{\Phi}_r\bar{\Phi}_r^T \alpha^{(j)} = \lambda_j \alpha^{(j)},
 \end{align}
which is solved for nonzero eigenvalues using Lemma~\ref{lowrankeig}.  
We choose $d \leq r$ kernel principal components, where the $j$-th component of a data point $x \in \mathcal{X}$ can be extracted by the projection  
\begin{align}
 p_j(x) = \bar{\Phi}_r(x) \cdot v^{(j)} = \frac{1}{\sqrt{\lambda_j}} \sum_{i=1}^m \alpha_i^{(j)} \bar{\Phi}_r(x_i) \cdot \bar{\Phi}_r(x).
\end{align}
\end{defi}

For the projections onto kernel principal axes holds the following relation.
\begin{cor}\label{cor:kpcaproj}
Given data points $\mathbf{y} = \{y_1,\hdots,y_\ell\} \subseteq \mathcal{X}$ their $j$-th kernel principal components are collectively calculated by 
\begin{align}\label{kpca_proj}
 (p_j(y_1),\hdots,p_j(y_\ell)) = \Big( \tfrac{1}{\sqrt{\lambda_j}} \,{\alpha^{(j)}}^T \bar{\Phi}_r\luk{}{[\mathbf{x}]} \Big) \bar{\Phi}_r[\mathbf{y}]^T = L^{(j)}  \bar{\Phi}_r[\mathbf{y}]^T,\quad j=1,\hdots,d,
\end{align}
where $\bar{\Phi}_r[.]$ stands for $\Phi_r[.]$ centered w.r.t. the training data. 
Moreover we defined the vectors $L^{(j)} = \tfrac{1}{\sqrt{\lambda_j}} \,{\alpha^{(j)}}^T \bar{\Phi}_r  \in \mathbb{R}^r$ for $j=1,\hdots,d$.
According to Corollary~\ref{cor:phiy} the projections \eqref{kpca_proj} are exact under the assumption of Lemma~\ref{lem:lowrank} that $K_{r,r}$ has full rank $r$. 
\end{cor}

Note that ${L^{(j)}}^T = \bar{\Phi}_r^T \alpha^{(j)}/\sqrt{\lambda_j}$ can be directly extracted from the algorithm of the low-rank eigenvalue problem \eqref{eqn:eigkpca2} owing to $w$ in Lemma~\ref{lowrankeig} and the fact that 
\begin{align}
 (\bar{\Phi}_r^T\bar{\Phi}_r)(\bar{\Phi}_r^T \alpha^{(j)}) = \lambda_j (\bar{\Phi}_r^T \alpha^{(j)}).
\end{align}
Hence, the low-rank kPCA needs to store a matrix of unit eigenvectors $L = \luk{}{[{L^{(1)}}^T|\cdots| {L^{(d)}}^T]} \in \mathbb{R}^{r \times d}$.  
Only  $K[\mathbf{y},\mathbf{x}_r] \in \mathbb{R}^{\ell \times r}$ is newly computed for projections onto the kernel principal axes in the course of the computation of $ \bar{\Phi}_r[\mathbf{y}]$.

\subsection{Low-rank kernel ridge regression and pre-image computation}\label{sec:lowrank_preimage}
Once the kPCA model is established from the training set $\mathbf{x} = \{x_1,\hdots,x_m\} \subseteq \mathcal{X}$, one can compute the projections onto the principal axes of new data points 
$\mathbf{y} = \{y_1,\hdots,y_\ell\} \subseteq \mathcal{X}$ via \eqref{kpca_proj}. Denote these projections with $P_d \phi_k(y_i) \in \mathcal{F}_k
,\,i=1,\hdots,\ell$. We will also be interested in 
finding an approximate pre-image $z_i \in \mathcal{X}$ from $P_d\phi_k(y_i)$ by solving the pre-image problems
\begin{align}
 z_i = \arg\min_{y} \| \luk{}{\phi_k(y)} - P_d\phi_k(y_i)\|^2,\,i=1,\hdots,\ell.
\end{align}
\luk{}{
This can be done by learning a pre-image map $\Gamma:\, \mathcal{F}_k \rightarrow \mathcal{X}$ that approximates 
\begin{align}
 y_i \approx z_i = \Gamma P_d\phi_k(y_i), \, i=1,\hdots,\ell, 
\end{align}
e.g., by establishing a kRR-model in a supervised learning approach using the training set $\mathbf{x}$ and its kPCA projections $P_d\phi_k(x_i),\,i=1,\hdots,m$ \cite{bakir2004learning}. 
We define the kRR problem \cite{shalev2014understanding, welling2013kernel} for determining the linear map $W$ representing approximately $\Gamma:\, \mathcal{F}_k \rightarrow \mathcal{X}$ in the form
\begin{align}\label{eqn:krr_preimage}
 \min_W \frac{1}{2} \sum_{i=1}^m \|x_i - W \cdot \phi_k \big(P_d(\Phi_r(x_i))\big) \|^2 + \frac{\alpha}{2} \|W\|^2,
\end{align}
with the regularization parameter \luk{}{$\alpha > 0$}.
Let us denote the kPCA projections of $\mathbf{x}$ and $\mathbf{y}$ with $P_d\Phi_r[\mathbf{x}] \in \mathbb{R}^{m \times d}$ and  $P_d\Phi_r[\mathbf{y}] \in \mathbb{R}^{\ell \times d}$, respectively.  
The (dual) solution to \eqref{eqn:krr_preimage} takes the form 
\begin{align}\label{eqn:dualsolkrr}
 W^T 
 = \phi_k(P_d\Phi_r[\mathbf{x}])^T \cdot \big(K[P_d\Phi_r[\mathbf{x}]] + \alpha I\big)^{-1} X,
\end{align} 
}
where $X = [x_1|\cdots|x_m]^T \in \mathbb{R}^{m \times N}$ and $N = \textrm{dim}(\mathcal{X})$ assumed to be finite here. 
\luk{}{The map $\phi_k$ is assumed to act on the rows of $P_d\Phi_r[\mathbf{x}]$ with $\phi_k(P_d\Phi_r[\mathbf{x}])$ being of size $m \times \textrm{dim}(\mathcal{F}_k)$. 
 $W^T$ in \eqref{eqn:dualsolkrr} is of size $\textrm{dim}(\mathcal{F}_k) \times \textrm{dim}(\mathcal{X})$.}
The kRR predictions for the pre-images of the projections $P_d\Phi_r[\mathbf{y}] \in \mathbb{R}^{\ell \times d}$ are given as  
\begin{align}
 Z = \phi_k(P_d\Phi_r[\mathbf{y}]) \cdot W^T = K[P_d\Phi_r[\mathbf{y}],P_d\Phi_r[\mathbf{x}]]\, B \in \mathbb{R}^{\ell \times N},\, B = (K[P_d\Phi_r[\mathbf{x}]] + \alpha I)^{-1} X \in \mathbb{R}^{m \times N}.
\end{align}
By utilizing the low-rank kernel approach from the previous Sec.~\ref{sec:lowrankkpca} the result for the pre-image prediction gets
\begin{align}\label{eqn:preimage}
 Z = \hat{\Phi}_r[\mathbf{y}]\,\luk{}{\big(}\hat{\Phi}_r[\mathbf{x}]^T B\luk{}{\big)} \in \mathbb{R}^{\ell \times N},\quad B = (\hat{\Phi}_r[\mathbf{x}]\, \hat{\Phi}_r[\mathbf{x}]^T + \alpha I)^{-1} X \in \mathbb{R}^{m \times N},
\end{align}
where the low-rank approximations $K[P_d\Phi_r[\mathbf{x}]] \approx \hat{\Phi}_r[\mathbf{x}]\,\hat{\Phi}_r[\mathbf{x}]^T$ and 
\luk{}{$K[P_d\Phi_r[\mathbf{y}],P_d\Phi_r[\mathbf{x}]] \approx \hat{\Phi}_r[\mathbf{y}]\, \hat{\Phi}_r[\mathbf{x}]^T$} are used with $\hat{\Phi}_r[\mathbf{x}]\in \mathbb{R}^{m \times r}, $\luk{}{$\, m \geq r,$ computed by 
Lemma~\ref{lem:lowrank}} and $\hat{\Phi}_r[\mathbf{y}] \in \mathbb{R}^{\ell \times r}$ \luk{}{ by Cor.~\ref{cor:phiy}}. Note that only the matrix $\hat{\Phi}_r[\mathbf{x}]^T B \in \mathbb{R}^{r \times N}$ has to be stored. 
The inverse in \eqref{eqn:preimage} can be expressed via the Sherman-Morrison-Woodbury (SMW) formula, i.e.,
\begin{align}\label{SMW}
 (\hat{\Phi}_r[\mathbf{x}]\, \hat{\Phi}_r[\mathbf{x}]^T + \alpha I)^{-1} = \alpha^{-1} I - \alpha^{-2} \hat{\Phi}_r[\mathbf{x}] \, (I + \alpha^{-1} \hat{\Phi}_r[\mathbf{x}]^T \hat{\Phi}_r[\mathbf{x}])^{-1}  \,\hat{\Phi}_r[\mathbf{x}]^T,
\end{align}
which only requires to solve linear systems of size $r \times r$ instead of $m \times m$. \luk{}{Alternatively, one can use the "push-through identity", i.e., $A(BA+\alpha I )^{-1} = (AB+\alpha I)^{-1}A$ for appropriately sized matrices $A$ and $B$ (simple proof by multiplication with the respective inverses on the right and left hand side) to arrive from \eqref{eqn:preimage} at
\begin{align}\label{eqn:preimage2}
 Z = \hat{\Phi}_r[\mathbf{y}]\, (\hat{\Phi}_r[\mathbf{x}]^T\, \hat{\Phi}_r[\mathbf{x}] + \alpha I)^{-1}\hat{\Phi}_r[\mathbf{x}]^T X \in \mathbb{R}^{\ell \times N}.
\end{align}}

In the course of the later prediction of micromagnetic time-evolution we will also use this low-rank version of kRR to estimate the time-stepping maps in feature space. 
\luk{}{In general, if we want to model a dependency of input data $\mathbf{x} = \{x_1,\hdots,x_m\} \subseteq \mathcal{X}$ and 
output data $\tilde{\mathbf{x}} = \{\tilde{x}_1,\hdots,\tilde{x}_m\} \subseteq \mathcal{Y}$ by a linear map $W:\, \mathcal{F}_k \rightarrow \mathcal{Y}$, the related kRR problem is  
\begin{align}\label{eqn:krr_gen}
 \min_W \frac{1}{2} \sum_{i=1}^m \|\tilde{x}_i - W\cdot\phi_k (x_i) \|^2 + \frac{\alpha}{2} \|W\|^2, \quad \alpha >0.
\end{align}
The (dual) solution of \eqref{eqn:krr_gen} is 
\begin{align}\label{eqn:dualsolkrr2}
 W^T = \phi_k(X)^T\cdot \big(\phi_k(X)\cdot \phi_k(X)^T + \alpha I \big)^{-1} \tilde{X} = \phi_k(X)^T\cdot \big(K[\mathbf{x}] + \alpha I \big)^{-1} \tilde{X},
\end{align}
where the data $\mathbf{x}$ and $\tilde{\mathbf{x}}$ are assembled into arrays $X$ and $\tilde{X}$ of shape $m \times \textrm{dim}(\mathcal{X})$ and $m \times \textrm{dim}(\mathcal{Y})$, respectively, 
and $\phi_k$ is assumed to act on 
the rows of $X$ with $\phi_k(X)$ being of size $m \times \textrm{dim}(\mathcal{F}_k)$.  $W^T$ in \eqref{eqn:dualsolkrr2} is of size $\textrm{dim}(\mathcal{F}_k) \times \textrm{dim}(\mathcal{Y})$. 
The low-rank version is established in an analogue way as above, reducing the solution operator $W$ to size $\textrm{dim}(\mathcal{Y}) \times r$.}

\subsection{Numerical validation of the low-rank kPCA}

We summarize the low-rank kPCA and pre-image procedure in algorithm~\ref{alg:low-rank-kPCA}. 
This generates the unit norm eigenvectors $L^{(j)},\, j=1,\hdots,d$ for the prediction of new data according to \eqref{kpca_proj} as well as 
the operator for the pre-image map $\hat{\Phi}_r[\mathbf{x}]^T B$ in \eqref{eqn:preimage}.\\ 

\begin{algorithm}\caption{Low-rank kPCA and pre-image}\label{alg:low-rank-kPCA}
\textbf{Data:}{\quad Training data $\mathbf{x}=\luk{}{\{}\mathbf{x}_r,\mathbf{x}_{m-r}\luk{}{\}} \subseteq \mathcal{X}$, kernel $k(.,.)$, $d \leq r$, $\alpha >0$. }\\
\textbf{Result:}{\quad \luk{}{Projection eigenvector matrix $L \in \mathbb{R}^{r \times d}$ from Cor.~\ref{cor:kpcaproj}}, truncated kernel PC's $P_d \Phi_r[\mathbf{x}]$, 
Operator for pre-image map $\hat{\Phi}_r[\mathbf{x}]^T B$ in \eqref{eqn:preimage}.}\\[0.2cm]
\textbf{Low-rank kPCA:}
\begin{itemize}
 \item Calculate $\Phi_r$ in \eqref{eqn:phi}.
 \item Solve low-rank eigenvalue problem \eqref{eqn:eigkpca2} for $d \leq r$ eigenvectors of unit length.  
 \end{itemize}
 \textbf{Low-rank pre-image map:}  
 \begin{itemize}
  \item Calculate  $K[P_d\Phi_r[\mathbf{x}]] \approx \hat{\Phi}_r[\mathbf{x}]\,\hat{\Phi}_r[\mathbf{x}]^T$.
  \item Calculate $\hat{\Phi}_r[\mathbf{x}]^T B$ in \eqref{eqn:preimage} using the SMW formula \eqref{SMW} \luk{}{or \eqref{eqn:preimage2}}.
 \end{itemize}
\end{algorithm}

 \begin{figure}[hbtp]
\centering 
\includegraphics[scale=.7]{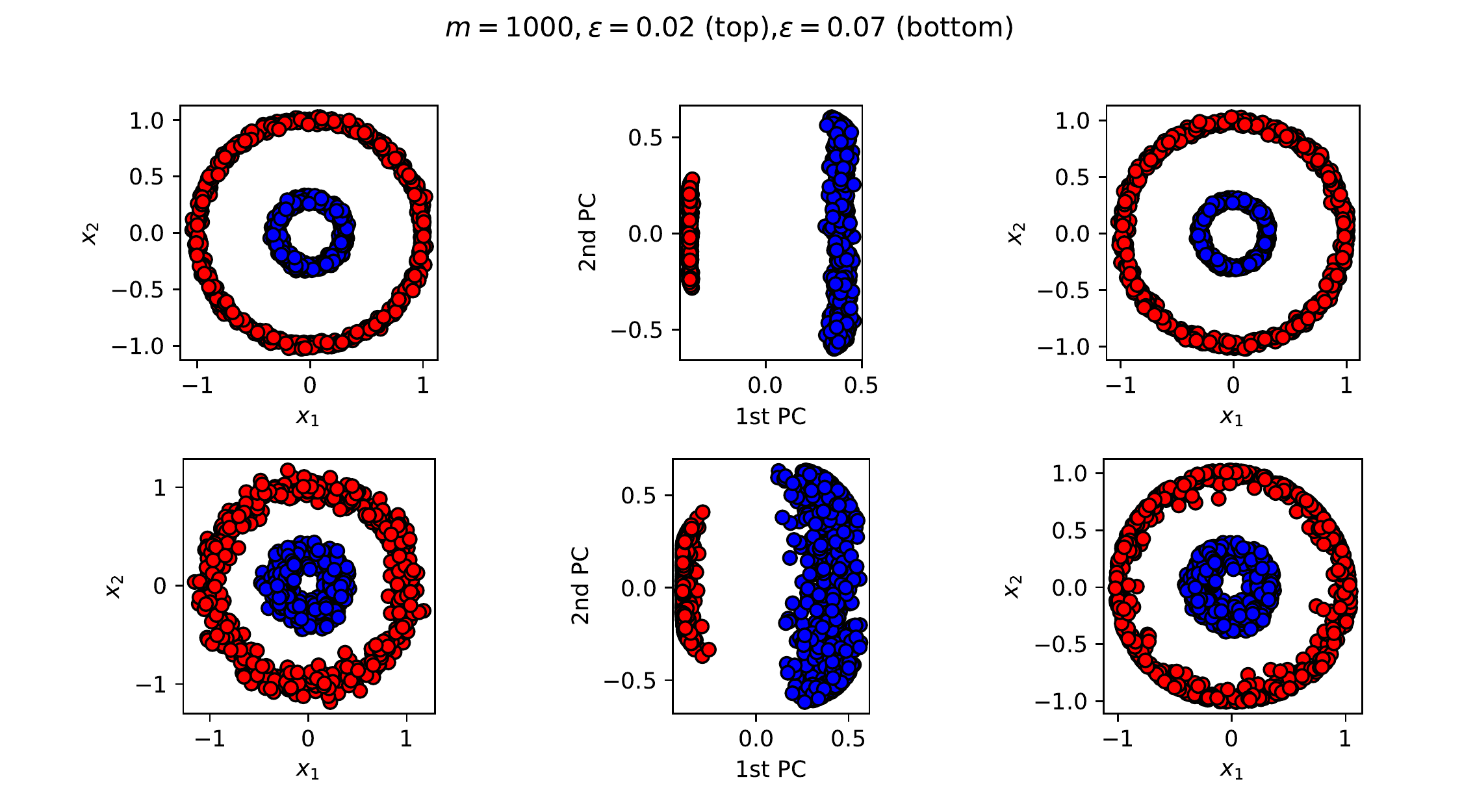}    
\caption{The original data set (left column), the kPCA transformed samples (middle column) and the pre-images (right column). 
Noise level $\varepsilon= 0.02$ (top) and $\varepsilon= 0.07$ (bottom). Number of training samples $m = 1000$. Rank $r=120$ is used.}\label{fig:kpca_ex}
\end{figure}
 \begin{figure}[hbtp]
\centering 
\includegraphics[scale=.7]{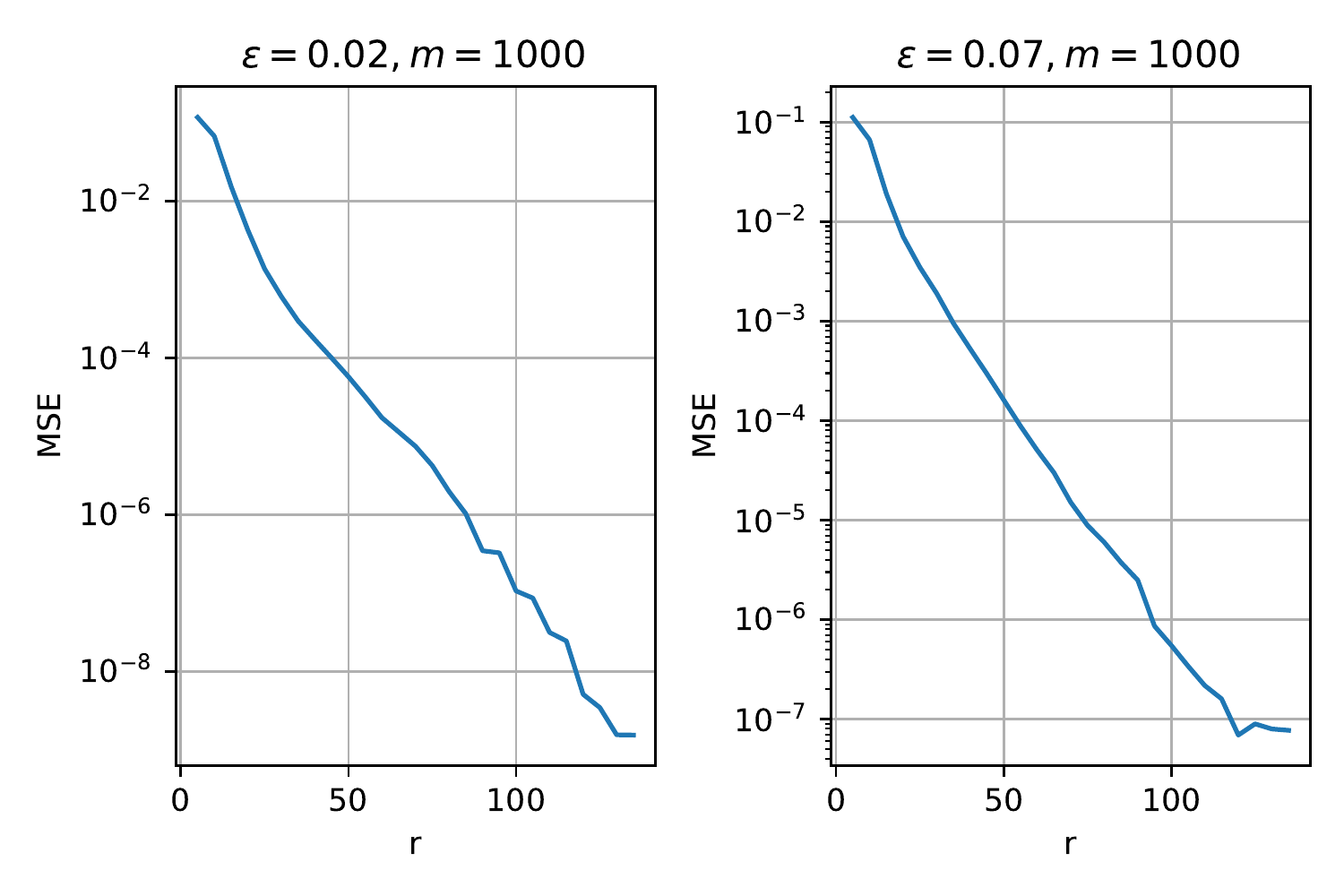}    
\caption{Mean squared error (MSE) for varying rank $r$ in the case of noise level $\varepsilon = 0.02$ and $\varepsilon = 0.07$.}\label{fig:mse_ex}
\end{figure}
 \begin{figure}[hbtp]
\centering 
\includegraphics[scale=.85]{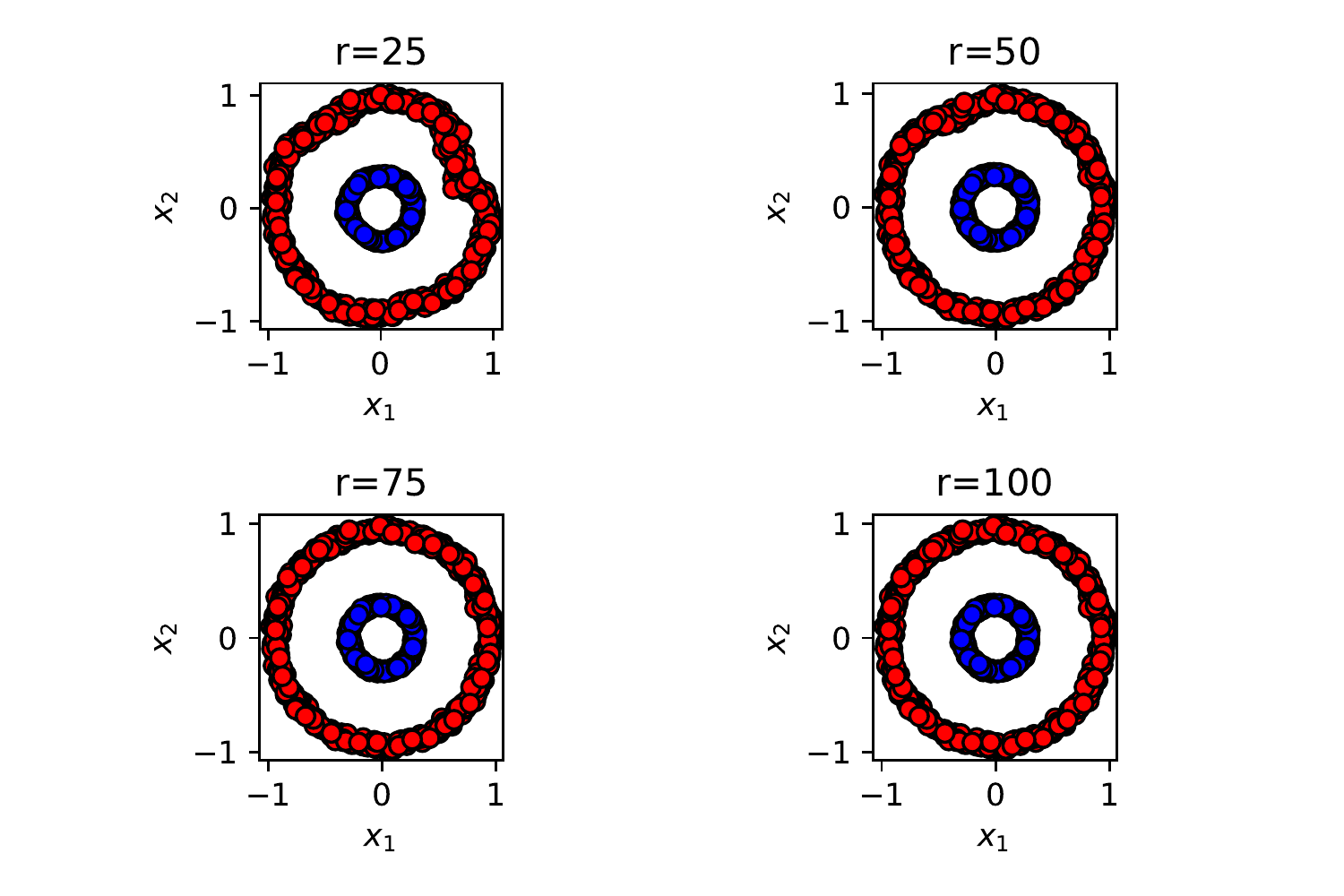}    
\caption{Pre-images for increasing rank $r$ and $\epsilon = 0.02$.}\label{fig:invtr1}
\end{figure}
 \begin{figure}[hbtp]
\centering 
\includegraphics[scale=.85]{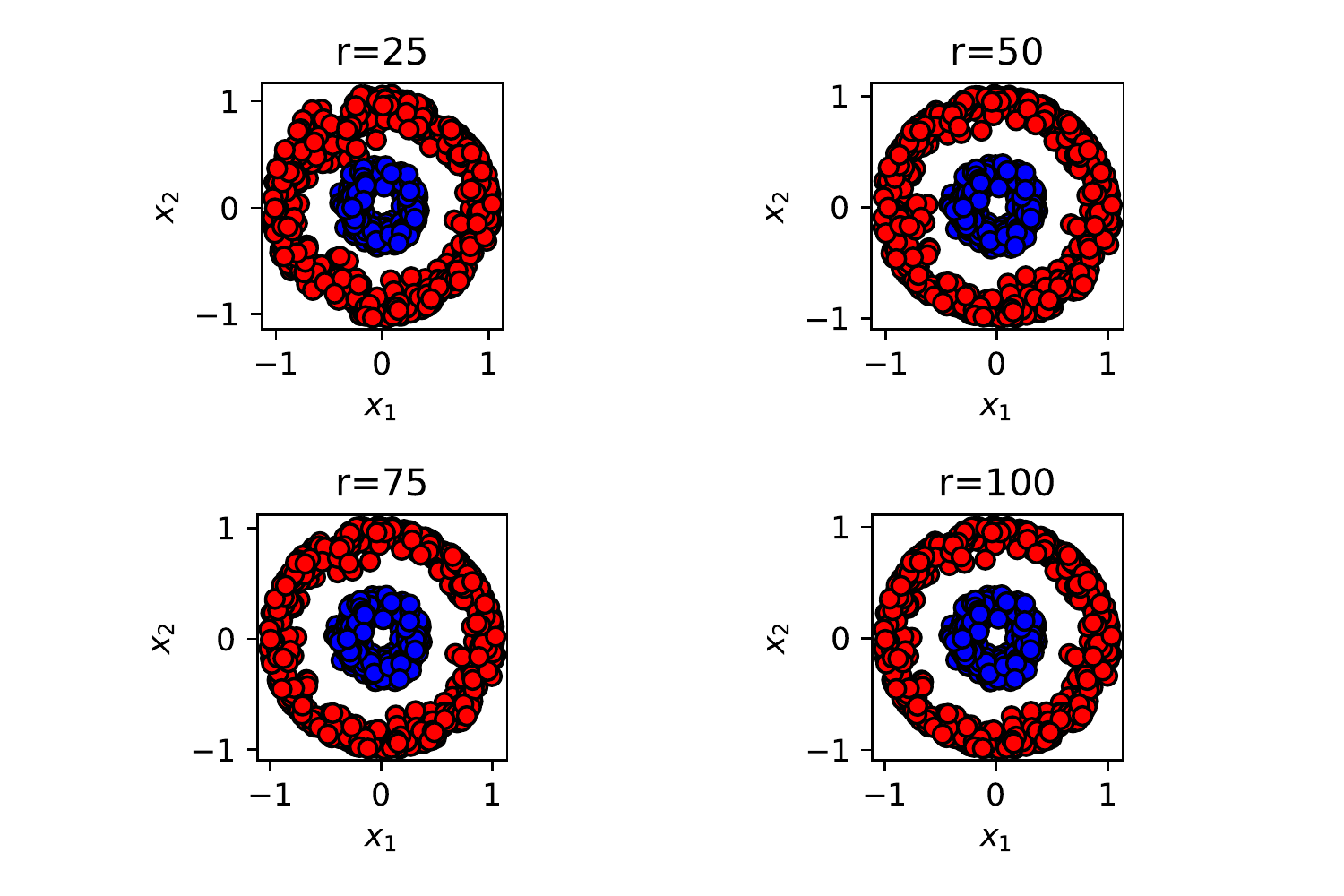}    
\caption{Pre-images for increasing rank $r$ and $\epsilon = 0.07$.}\label{fig:invtr2}
\end{figure}
\luk{}{The low-rank kPCA was implemented as an extension in the scikit learn Python software  \cite{scikitlearn}.}
We validate the low-rank version of kPCA and the pre-image solution via a test example from the scikit learn documentation, which uses both $m=1000$ training data and test data drawn from concentric circles with noise, see Fig.~\ref{fig:kpca_ex}. \luk{}{We use a RBF kernel with $\gamma=4$ and resolve three kernel principal components.}
Two noise levels $\varepsilon = 0.02$ and $0.07$ are used, where in both cases fast (exponential) convergence for increasing rank $r$ can be observed, see Fig.~\ref{fig:mse_ex}  
which shows the mean squared error of the pre-images of the predictions compared with the original data with varying rank $r$ used in the low-rank kPCA. 
The Figs.~\ref{fig:invtr1} and \ref{fig:invtr2} show the pre-images for increasing rank in the two noise cases, respectively.
\luk{}{Table \ref{tab:lowrankkpca} shows the training and prediction times for varying $m$ and $r$ and compares the low-rank kPCA components with those obtained from the dense (conventional) kPCA in terms of the mean squared error. We used a Intel(R) Core(TM) i7-4770K CPU \@ 3.50GHz. 
\begin{table}
\tabcolsep 6pt 
\caption{\luk{}{Cpu times in seconds for training and prediction of low-rank kPCA for varying samples $m$ and rank $r$. Numbers in brackets refer to computing times of the respective dense kPCA version. The last column shows the mean squared error of the 
computed three kernel principal components compared to the results obtained from the conventional kPCA.}}\label{tab:lowrankkpca}
\begin{center}
\begin{tabular}{c c c c c}
    $m$ &    $r$ &  training time &  prediction time &  mse \\
\hline\hline
 8000 &  100 &  0.075 (2.813) &    0.053 (1.696) &  0.132 \\ \hline
 4000 &  100 &  0.043 (0.791) &    0.028 (0.440) &  0.132 \\ \hline
 2000 &  100 &  0.037 (0.246) &    0.014 (0.153) &  0.092 \\ \hline
 1000 &  100 &  0.010 (0.069) &    0.004 (0.047) &  0.081 \\ \hline
  500 &  100 &  0.018 (0.030) &    0.003 (0.013) &  0.041 \\ \hline\hline
 1000 &  800 &  0.307 &    0.056 &  1.14e-18 \\ \hline
 1000 &  400 &  0.119 &    0.027 &  1.10e-18 \\ \hline
 1000 &  200 &  0.039 &    0.021 &  5.76e-16 \\ \hline
 1000 &  100 &  0.010 &    0.004 &  0.081 \\ \hline
 1000 &   50 &  0.013 &    0.002 &  0.055 \\ \hline
\end{tabular}
\end{center}
\end{table}
It also shows the scaling of the dense kPCA with the sample size $m$.}

\subsection{Learning maps between feature space elements with truncated components}
\begin{figure}[hbtp]
\centering 
\includegraphics[scale=1.8]{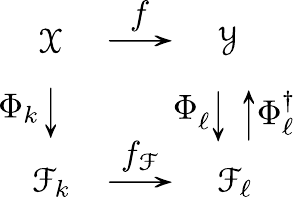}    
\caption{Illustration of the mappings in \eqref{eqn:mappings}.}\label{fig:mappings}
\end{figure}

\begin{figure}[hbtp]
\centering 
\includegraphics[scale=1.8]{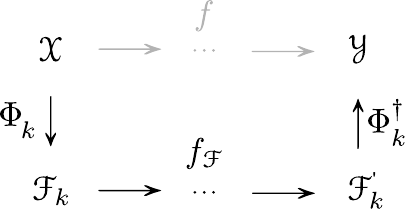}    
\caption{Illustration of the mappings involved in the feature space integration procedure.}\label{fig:featurespaceint}
\end{figure}

We denote $\mathcal{X}$ as the \textit{input set} and $\mathcal{Y}$ as the \textit{output set}. 
A general \textit{learning problem} is to estimate a map between inputs $x \in \mathcal{X}$ and outputs $y \in \mathcal{Y}$. 
The underlying mathematical task is that of estimating a map from an Hilbert space $\mathcal{V}$ by minimizing the risk functional 
\begin{align*}
 f^\ast \in \arg\min_{f \in \mathcal{V}} \mathcal{J}(f) := \int_{\mathcal{X}\times \mathcal{Y}} L(y,f(x))\, \textrm{d}\rho(x,y),
\end{align*}
on the measure space $\big(\mathcal{X} \times \mathcal{Y}, \Sigma_{\mathcal{X} \times \mathcal{Y}}, \rho \big)$ 
but with \textit{unknown} joint probability distribution $\rho$. If we have available inputs $x \in \mathcal{X}$ and outputs $y \in \mathcal{Y}$ from a given training set 
$(x_1,y_1), (x_2,y_2), \hdots, (x_m,y_m) \in \mathcal{X} \times \mathcal{Y},$ we can try to empirically solve the problem in a model class or hypothesis class like e.g. $\mathcal{H} = \{f(.;\alpha):\, \alpha \,\,\textrm{feasible parameter}\}$.
In \cite{exl2020learning} we defined $L$ as the distance in output feature space using 
a radial basis function as kernel $\ell:\,\mathcal{Y} \times \mathcal{Y} \rightarrow \mathbb{R}$ on the output set. This gives a RKHS $\mathcal{F}_\ell$ with associated map 
$\phi_\ell:\, \mathcal{Y} \rightarrow \mathcal{F}_\ell$ and $\ell(y,y^\prime) = \phi_\ell(y)\cdot \phi_\ell(y^\prime)$ and a loss expression $L(y,f(x)) = \|\phi_\ell(y) - \phi_\ell(f(x))\|_{\mathcal{F}_\ell}^2$, 
which can be expressed entirely through the kernel $\ell$ using the kernel trick. For the purpose of finding the minimizer $f^\ast$, only few kernel principal components of the representation of feature vectors are used and
a ridge regression is used in \cite{exl2020learning}. 
Generally, the problem of estimating the map $f$ can be decomposed in subtasks using the idea of \textit{kernel dependency estimation (KDE)} \cite{weston2003kernel}, where $f$ is the composition of three maps, i.e.,
\begin{align}\label{eqn:mappings}
 f = \phi_\ell^{\dagger} \circ f_{\mathcal{F}} \circ \phi_k, 
\end{align}
where $\phi_k:\,\mathcal{X} \rightarrow \mathcal{F}_k$ is the feature map for inputs associated with a kernel $k$, $f_{\mathcal{F}}:\,\mathcal{F}_k \rightarrow \mathcal{F}_\ell$ the map between input and output feature spaces and 
$\phi_\ell^\dagger:\,\mathcal{F}_\ell \rightarrow \mathcal{Y}$ an approximate inverse onto $\mathcal{Y}$ which is the \textit{pre-image map}, where here we will use the computational low-rank approach of section~\ref{sec:lowrank_preimage}. 
See Fig.~\ref{fig:mappings} for an illustration of the involved mappings. 
The process could be even established by circumventing the kPCA as described in \cite{CortesKDE}. To make up for the lack of a learned pre-image available from kPCA, such an approach would have to use a fixed point iteration for the pre-image map, which is possible as long as RBF kernels are used. However, a learned pre-image map is faster and more reliable. Furthermore this approach would not allow for a time-stepping procedure which entirely operates in feature space with reduced dimensionality. In the following method we rather estimate a map with (low-rank) kernel-ridge regression between truncated kPCA coordinate representations of elements in the actually infinite dimensional feature space.  
That is, we estimate a map between input and output representatives of the form $\big(p_1(x),p_2(x),\hdots,p_d(x)\big)$ and $\big(p_1(y),p_2(y),\hdots,p_d(y)\big)$, respectively, where we consider a truncated number 
of $d \in \mathbb{N}$ kernel principal components. We use kRR analogues to section~\ref{sec:lowrank_preimage}, where the matrix to be stored is of shape $r \times d$. The same kernel for the input and output space embedding is used, that is, $\mathcal{F}_k = \mathcal{F}_\ell$. 
This new approach drastically improves the quality of the prediction alongside with computational efficiency from the low-rank framework of section~\ref{sec:lowrankkpca} and  \ref{sec:lowrank_preimage}. Our learning approach works entirely in feature space, that is, all time steps are learned within the reduced dimensional setting and the pre-image is used after the final time step, see Fig.~\ref{fig:featurespaceint}, which illustrates the feature space integration scheme.

%

\section{Prediction of magnetization dynamics}\label{sec:numerics}

The mathematical description of magnetization dynamics in a magnetic body $\Omega \subset \mathbb{R}^3$ is through the \textit{Landau-Lifschitz-Gilbert} (LLG) equation \cite{kronmueller}. 
In micromagnetism we consider the magnetization as a vector field $\boldsymbol{M}(x,t) = M_s \boldsymbol{m}(x,t),\,|\boldsymbol{m}(x,t)| = 1$ 
depending on the position $x \in \Omega$ and the time $t\in \mathbb{R}$. The LLG equation is given in explicit form as  
\begin{align}\label{eqn:LLGM}
 \frac{\partial \boldsymbol{M}}{\partial t} &\, = -\frac{\gamma_0}{1+\alpha^2} \, \boldsymbol{M} \times \boldsymbol{H} - \frac{\alpha\,\gamma_0}{(1+\alpha^2) M_s} \, \boldsymbol{M} \times \big(\boldsymbol{M} \times \boldsymbol{H}\big),
\end{align}
where $\gamma_0$ is the gyromagnetic ratio, $\alpha$ the damping constant and $\boldsymbol{H}$ the effective field, which is the sum of nonlocal and local fields such as the stray field and the exchange field, respectively, and the external field $\boldsymbol{h}\in\mathbb{R}^3$ with length $h$. 
The stray field arises from the magnetostatic Maxwell equations, that is the whole space Poisson equation for the scalar potential $u_d$
\begin{align}
 \Delta u_d = \nabla\cdot \mathbf{M}\quad \textrm{in}\,\, \mathbb{R}^3, 
\end{align} with $\mathbf{H}_d = -\nabla u_d$. 
The exchange term is a continuous micro-model of Heisenberg exchange, that results in $\mathbf{H}_{ex} = \tfrac{2\,A}{\mu_0 M_s^2} \Delta \mathbf{M}$, where $\mu_0$ is the vacuum permeability, $M_s$ the saturation magnetization and $A$ the exchange constant.
Equation \eqref{eqn:LLGM} is a time-dependent partial differential equation in $3$ spatial dimensions supplemented with an initial condition $\boldsymbol{M}(x,t=0) = \boldsymbol{M}_0$ and (free) Neumann boundary conditions. For further details on micromagnetism the interested reader is referred to the literature \cite{brown1963micromagnetics,aharoni2000introduction,kronmueller}.
Typically, equation \eqref{eqn:LLGM} is numerically treated by a semi-discrete approach \cite{suess2002time,donahue1999oommf,d2005geometrical,exl2017extrapolated}, 
where spatial discretization by collocation using finite differences or finite elements leads to a rather large system of ordinary differential equations. Clearly, the evaluation of the right hand side of the system is very expensive mostly due to the stray field, 
hence, effective methods are of high interest. Our proposed \textit{data-driven} approach yields a predictor model for the magnetization dynamics without any need for  field evaluations after a data generation and training phase has been established as a pre-computation.

\subsection{Data structure for the time stepping learning method}
\begin{figure}[hbtp]
\centering 
\includegraphics[scale=0.7]{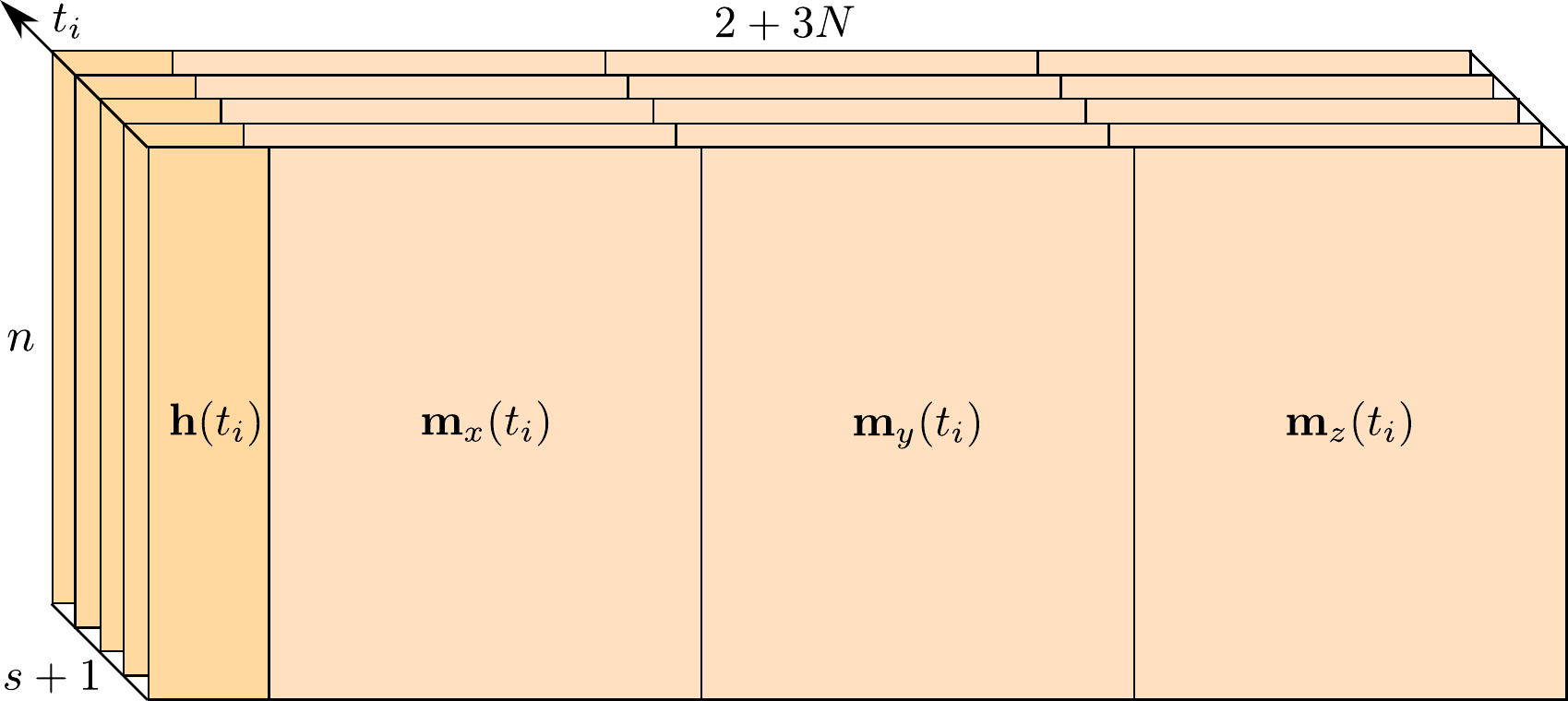}    
\caption{Data tensor $\bar{\mathcal{D}}$.}\label{fig:datatensor}
\end{figure}

Following \cite{exl2020learning} we generate data associated with the NIST $\mu$MAG Standard problem $\#4$ \cite{mumag4}. The geometry is a magnetic thin film of size $500 \times 125 \times 3$ nm$^3$ with material parameters of permalloy: 
$A = 1.3 \times 10^{-11}$ J/m, $M_s = 8.0 \times 10^5$ A/m, $\alpha = 0.02$ \luk{}{and $\gamma_0 = 2.211 \times 10^5$ m/(As)}. The initial state is an equilibrium s-state, obtained after applying and slowly reducing a saturating field along the diagonal direction $[1,1,1]$ to zero. 
Then two scenarios of different external fields are studied: field $1$ of magnitude $25$mT is applied with an angle of $170^\circ$ c.c.w. from the positive $x$ axis, field $2$ of magnitude $36$mT is applied with an angle of $190^\circ$ c.c.w. from the positive $x$ axis. 
For data generation we use a spatial discretization of $100 \times 25 \times 1$ and apply finite differences \cite{miltat2007numerical} to obtain a system of ODEs that is then solved with a projected Runge-Kutta method of second order with constant step size of $40$fs.\\
We denote the number of discretization cells with $N$. For the purpose of collecting training data samples we use numerically obtained approximations for $n\in \mathbb{N}$ different external field values. Following the splitting of training data in \cite{exl2020learning} the external field is either in the range of the \textit{field $1$}  
\begin{align}\label{eqn:hext1}
 \mathbf{H}_{ext,1}:\, \|\mathbf{H}_{ext,1}\| =: h \in [20,30]\textrm{mT},\, \arg \mathbf{H}_{ext,1} =: \varphi \in [160^\circ,180^\circ]
\end{align}
 or  in the range of the \textit{field $2$} 
\begin{align}\label{eqn:hext2}
 \mathbf{H}_{ext,2}:\, \|\mathbf{H}_{ext,2}\| =: h \in [30,40]\textrm{mT},\, \arg \mathbf{H}_{ext,2} =: \varphi \in [180^\circ,200^\circ].
\end{align} 
We use $n = 300$ for each data set, which, however, will be effectively reduced to a rank $r \leq n$ by the later low-rank approach. For $s=100$ time steps we 
assemble the data into a $3$-tensor $\mathcal{D}$, respectively $\bar{\mathcal{D}}$, defined slice-wise by
\begin{align}
 \mathcal{D} \in \mathbb{R}^{(s+1)\times n \times 3N}:\, \mathcal{D}(i,:,:) = [\mathbf{m}_x(t_i)|\mathbf{m}_y(t_i)|\mathbf{m}_z(t_i)] \in \mathbb{R}^{n \times 3N},\, i = 0,\hdots,s,
\end{align}
and
\begin{align}
 \bar{\mathcal{D}} \in \mathbb{R}^{(s+1)\times n \times (3N+2)}:\, \bar{\mathcal{D}}(i,:,:) = [\mathbf{h}(t_i)|\mathbf{m}_x(t_i)|\mathbf{m}_y(t_i)|\mathbf{m}_z(t_i)] \in \mathbb{R}^{n \times (3N+2)},\, i = 0,\hdots,s,
\end{align}
where $\mathbf{m}_q(t_i) \in \mathbb{R}^{n \times N},\, q=x,y,z$ denotes the magnetization component grid vector at time $t_i$ for each of the $n$ field values and $\mathbf{h}(t_i) \in \mathbb{R}^{n \times 2}$ consists of the external field samples at time $t_i$ with $h$ and $\varphi$ component each. Fig.~\ref{fig:datatensor} illustrates the data tensor $\bar{\mathcal{D}}$, which equals $\mathcal{D}$ extended by the external field values.

Selection of basis vectors for the low-rank procedure (compare with $\mathbf{x}_r$ in section~\ref{sec:methods}) is accomplished by choosing $r$ field values and collecting  the corresponding discrete magnetization trajectories for all $s+1$ time points for each chosen field value. This results in a reduced sample size of $(s+1)r \leq (s+1)n = m$. 

\begin{figure}[hbtp]
\centering 
\includegraphics[scale=0.7]{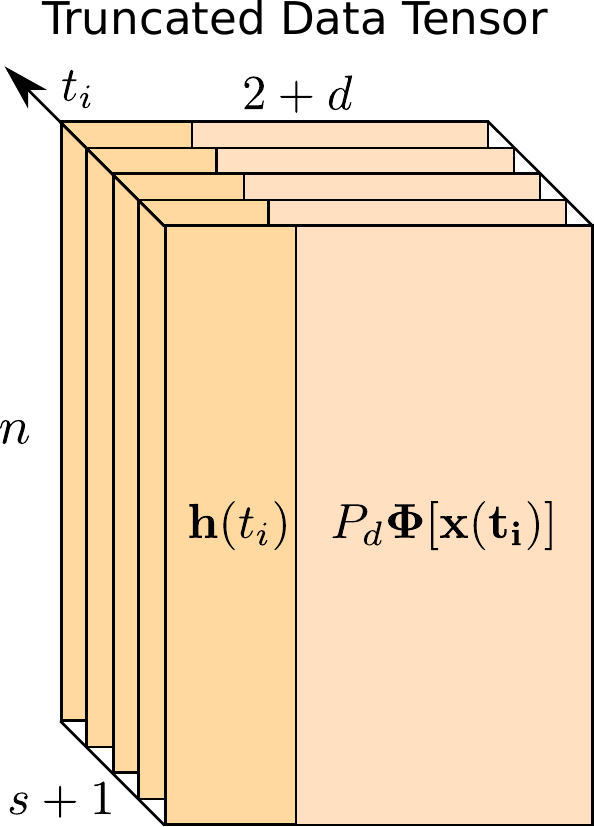}    
\caption{Illustration of the truncated low-rank kPCA data tensor $\bar{\mathcal{D}}_{\mathcal{F}}$.}\label{fig:compression}
\end{figure}

\begin{figure}[hbtp]
\centering 
\includegraphics[scale=0.7]{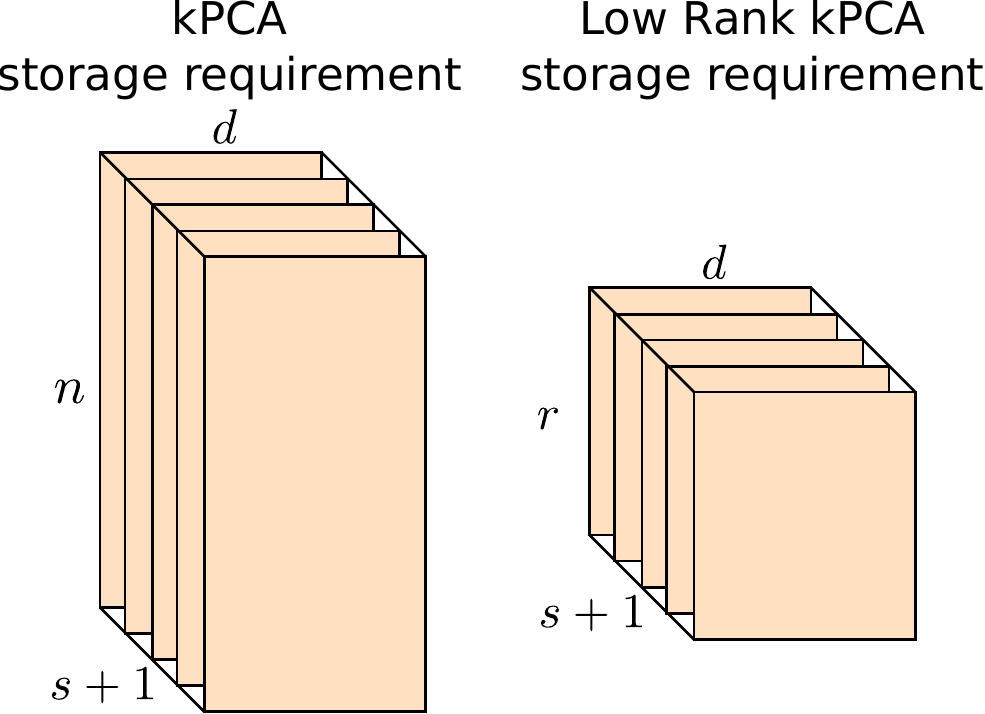}    
\caption{Illustration of the storage requirements in the low-rank kPCA and the low-rank kRR for the time stepping model (right), compared with full-rank kPCA (left). Storage of the compressed tensor is only $\mathcal{O}((s+1)dr)$.}\label{fig:storagetensor}
\end{figure}

In the course of the time-stepping learning via low-rank kPCA the data tensor is used with reduced dimensionality. Note that we have $d \leq r(s+1) \leq n(s+1) = m$. We denote the reduced dimensional data tensor resulting from the low-rank kPCA approach with $\mathcal{D}_{\mathcal{F}} \in \mathbb{R}^{(s+1)\times n \times d}$. 
Fig.~\ref{fig:compression}  shows the compressed (resp. truncated) data tensor $\bar{\mathcal{D}}_{\mathcal{F}}$, where the large grid size $3N$ is reduced to $d$ and the field is appended, compare with the original data tensor from Fig.~\ref{fig:datatensor}. Additionally we illustrate in Fig.~\ref{fig:storagetensor} the tensor required in storage to project new data onto the kernel principal components, as well as, involved in the kRR to fit the time stepping maps. 

Time stepping maps are now learned by taking reduced dimensional kPCA input and output data tensor to fit a kRR model. 
In its simplest form one can use a one step scheme mapping from $t \rightarrow t + \Delta t$ by taking input data $\mathcal{D}_{\mathcal{F}}^{(0)} \in \mathbb{R}^{s\times n \times d}$ defined via the slices 
$\mathcal{D}_{\mathcal{F}}(i,:,:),\,i=0,\hdots,s-1$, and output data $\mathcal{D}_{\mathcal{F}}^{(1)} \in \mathbb{R}^{s\times n \times d}$ 
defined via the slices $\mathcal{D}_{\mathcal{F}}(i,:,:),\,i=1,\hdots,s$, which corresponds to data shifted by one time step $\Delta t$. However, inspired by \cite{kovacs2019learning}, we found enhanced stability by introducing time stepping with multi-steps, e.g., choosing $\nu$ steps in a scheme $\{t,t+\Delta t,\hdots,t+(\nu-1)\Delta t\}\rightarrow t+\nu \Delta t$. For that purpose we choose a time stepping number $\nu (<s) \in \mathbb{N}$ and take the following training input and output sets:  
\begin{align}
\textrm{input:\,}\quad \{\bar{\mathcal{D}}_{\mathcal{F}}^{(0)},\hdots,\bar{\mathcal{D}}_{\mathcal{F}}^{(\nu-2)},\bar{\mathcal{D}}_{\mathcal{F}}^{(\nu-1)} \}, \quad \textrm{output:} \quad \{\mathcal{D}_{\mathcal{F}}^{(\nu)} \},    
\end{align}
where $\bar{\mathcal{D}}_{\mathcal{F}}^{(\nu-j)}\in \mathbb{R}^{(s-\nu)\times n \times (d+2)},\, j \in \{1,\hdots,\nu\}$ are defined via the slices $\bar{\mathcal{D}}_{\mathcal{F}}(i,:,:),\,i= \nu-j\hdots,s-j$ and $\mathcal{D}_{\mathcal{F}}^{(\nu)} \in \mathbb{R}^{(s-\nu)\times n \times d}$ via slices $\mathcal{D}_{\mathcal{F}}(i,:,:),\,i= \nu,\hdots,s$. \\

\luk{}{Besides the storage requirements for the tensor in Fig.~\ref{fig:storagetensor} the low-rank approaches need to store the realizations of the feature map $\Phi_r[\mathbf{x}]$ which are of size $n(s+1) \times r(s+1)$, since $m = n(s+1)$ and the rank $r$ from Sec.~\ref{sec:lowrankkpca} is a multiple of the number of time points $s+1$. Essentially, overall computational costs and storage requirements improve due to the smaller kernel matrix for the low-rank compared to dense versions. In detail, computation of the low-rank approximation to the kernel matrix for the magnetization data costs $n(s+1)^2r+r^2(s+1)^2$ kernel function evaluations, a cost of $\mathcal{O}\big(r^3s^3\big)$ for the root and $\mathcal{O}\big(ns^3r^2\big)$ for the matrix multiplication. Storage of $\Phi_r[\mathbf{x}]$ amounts to $n(s+1)^2r$. Learning effort for the kPCA is dominated by the computation of the co-variance matrix $\Phi_r[\mathbf{x}]^T\Phi_r[\mathbf{x}]$ and the eigenvalue decomposition, scaling $\mathcal{O}\big(s^3r^2n\big)$ and $\mathcal{O}\big(r^3s^3\big)$, respectively. The kPCA projection and its storage is cheap, both scaling only $\mathcal{O}\big(d sr\big)$. Storage for the pre-image projection $\Phi_r[\mathbf{x}]^T B$ is $r(s+1) \times 3N$, where the training phase costs $\mathcal{O}\big(s^3r^3+s^3r^2n\big) + \mathcal{O}\big(rns^2 N\big)$. Computational cost for one pre-image computation is $\mathcal{O}\big(r s N\big)$. For the feature space integration we need to fit $n(s-\nu)$ samples and targets with $\nu(d+2)$ and $d$ components, respectively, using low-rank kRR.  The parameter $\nu$ enters linearly into the scaling for the computation of the kernel matrix in the low-rank approximation. The training phase computation for kRR needs  $\mathcal{O}\big(s^3r^3+s^3r^2n\big) + \mathcal{O}\big(rns^2 d\big)$ operations and storage amounts here 
to $r(s+1)d$. Computational cost for one time-step is $\mathcal{O}(rsd)$.
} 

\subsection{Numerical experiments}
The data generation and cross-validation were performed using the Vienna Scientific Cluster (VSC). We used the Python machine learning package scikit learn \cite{scikitlearn} which we extended by the low-rank kPCA variant with pre-image computation and low-rank kRR introduced in section~\ref{sec:methods} above. 
We divide the numerical experiments into two categories. First we focus on the important validation of model and method specific hyper-parameters such as the kernel defining $\gamma > 0$, the time stepping number $\nu \in \mathbb{N}$ and the number of kernel principal components $d \in \mathbb{N}$. Afterwards we apply the low-rank method to the micromagnetic benchmark and study the dependence on the rank $r \in \mathbb{N}$.  \luk{}{As described in the previous section we take $n=300$.} 

\begin{figure}[hbtp]
\centering 
\includegraphics[scale=0.8]{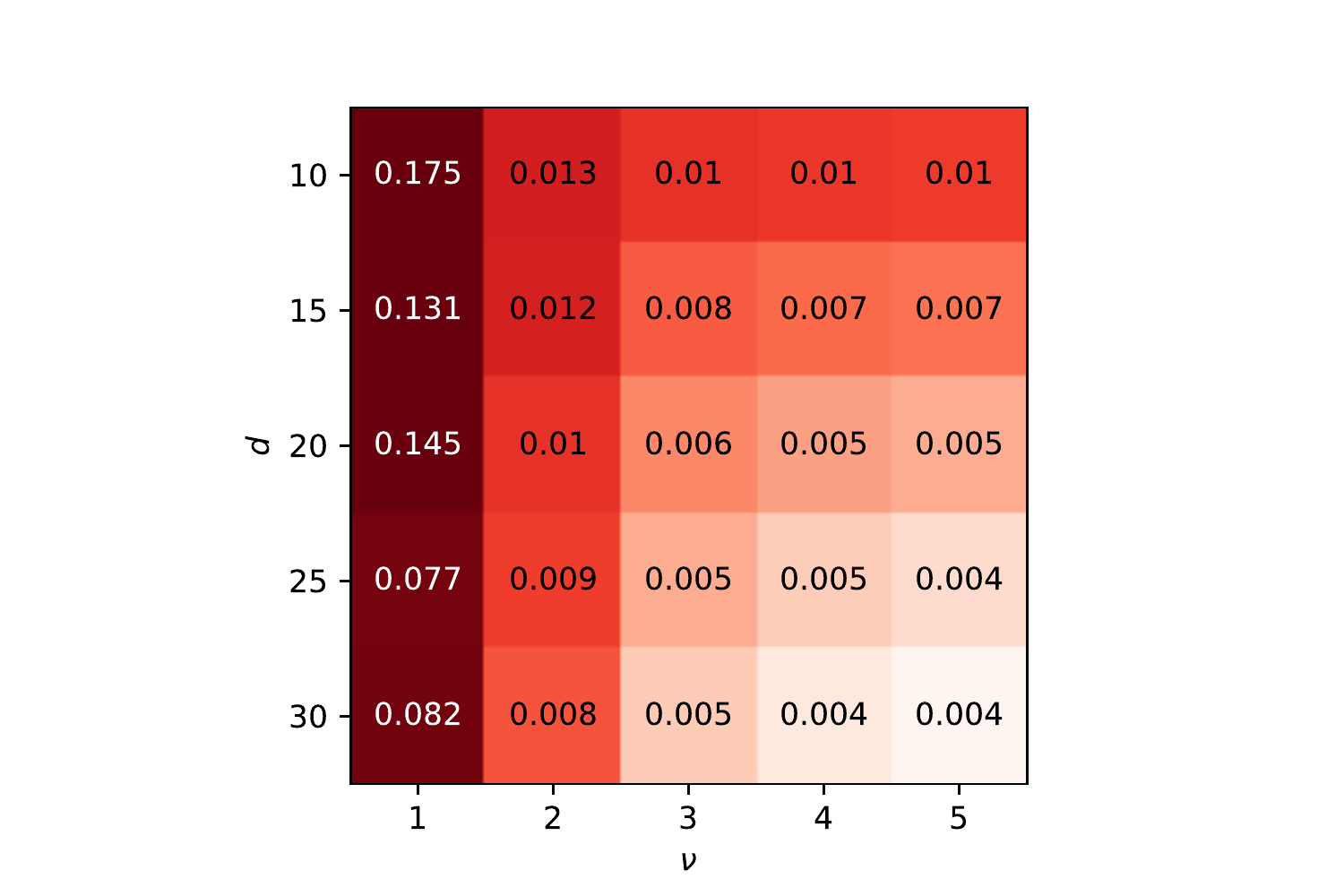}    
\caption{Cross validation table. Mean error norm in the magnetization for the prediction of $\mathbf{H}_{ext,1}$-data for varying number of kernel principal components $d$ and step parameter $\nu$.}\label{fig:heatmap1}
\end{figure}

\begin{figure}[hbtp]
\centering 
\includegraphics[scale=0.8]{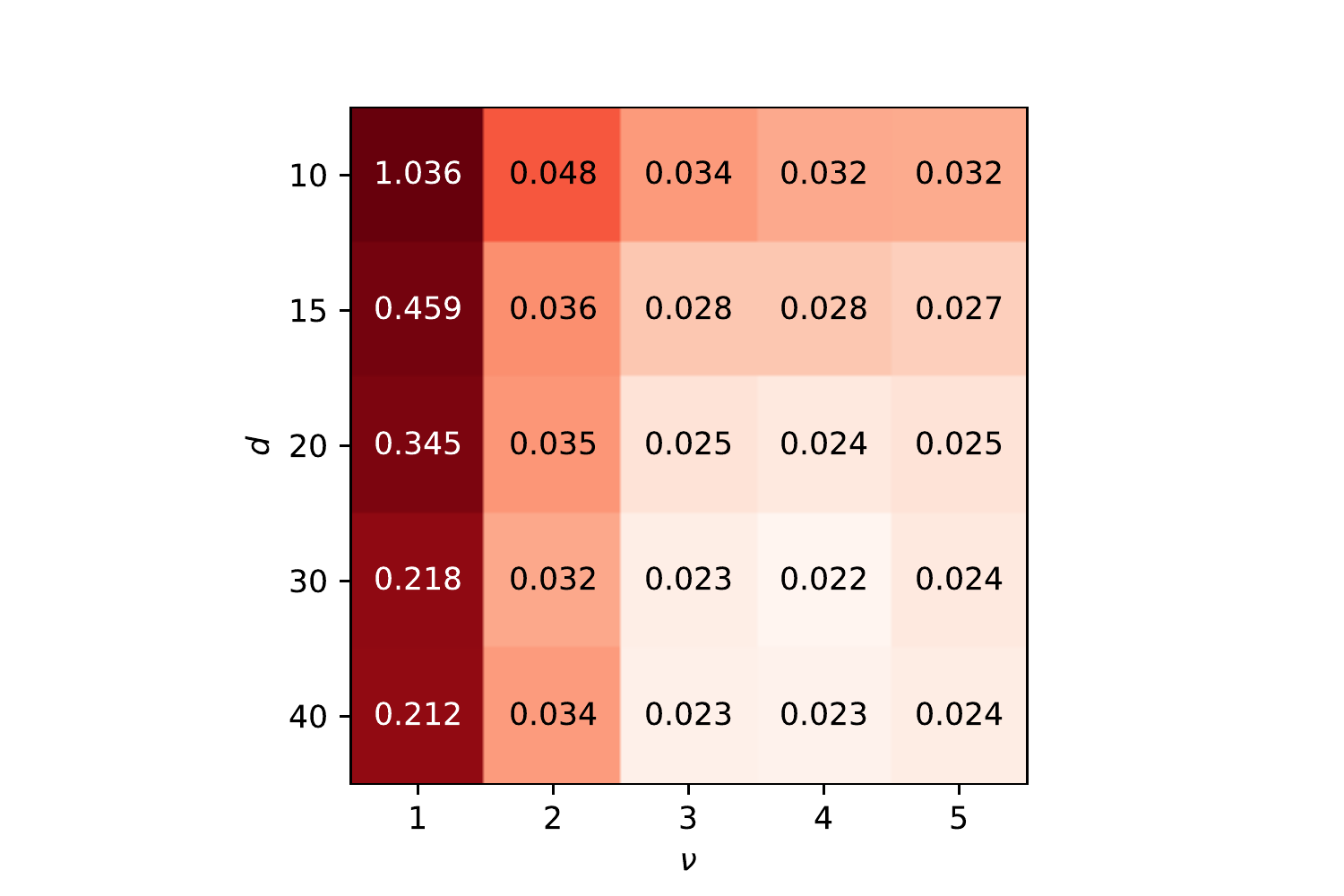}    
\caption{Cross validation table. Mean error norm in the magnetization for the prediction of $\mathbf{H}_{ext,2}$-data for varying number of kernel principal components $d$ and step parameter $\nu$.}\label{fig:heatmap2}
\end{figure}

\vspace*{0.5cm}
\textbf{Cross-validation of the hyper-parameters.}\\
We determine the hyper-parameters $\gamma, \nu$ and $d$ via grid search. For that purpose we measure 
the mean error norms in the magnetization between the prediction and the simulation of $1$ns for the standard problem in both ranges of field $1$ and $2$. This shows that a (default) value of $\gamma = 1/\luk{}{(3N)}$ is quite optimal. Furthermore, the regularization parameters in the kRR were chosen to be between $0.001$ and $0.01$ \luk{}{and $\gamma_{kRR} = 1$ performed sufficiently well}.  
Fig.~\ref{fig:heatmap1} and Fig.~\ref{fig:heatmap2} show for varying $d$ and $\nu$ the mean error norms in the magnetization between \luk{}{all} the predictions and simulations of $1$ns for the standard problem in the range of field $1$ and $2$ (compare with \eqref{eqn:hext1} and \eqref{eqn:hext2}), respectively, obtained from a $10$-fold cross-validation with random split strategy and $10\%$ test size. Here we used a rather large rank $r=40$.

\begin{figure}[hbtp]
\centering 
\includegraphics[scale=0.70]{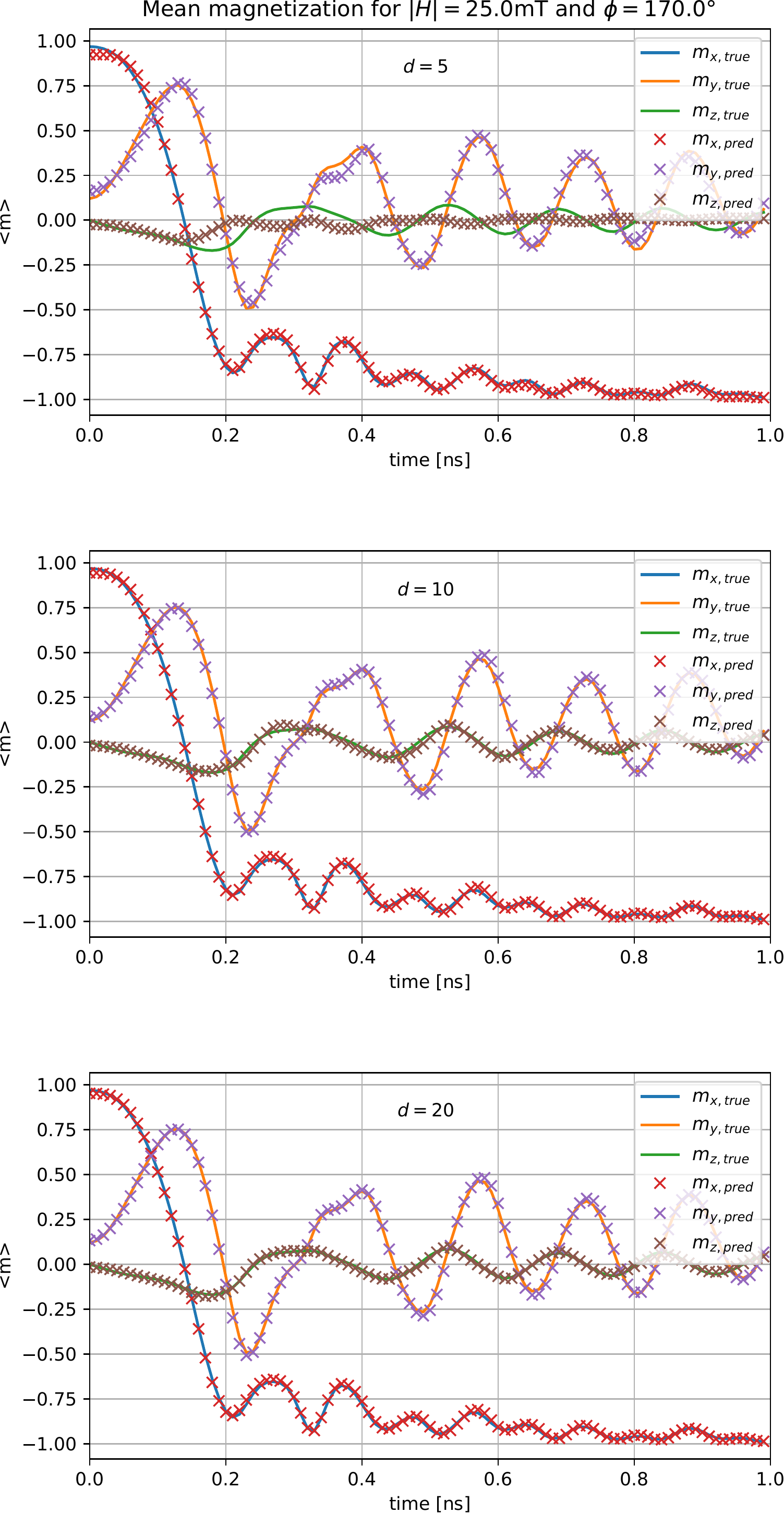}    
\caption{Predictions versus computed results for mean magnetization in the field $1$ case for varying number of kernel principal components $d = 5,10$ and $20$. The parameters were chosen as follows: kernel parameter $\gamma = 1/\luk{}{(3N)}$ and time-stepping $\nu=3$. A rank of $r=30$ was used.}\label{fig:field1_varying_d}
\end{figure}

\begin{figure}[hbtp]
\centering 
\includegraphics[scale=0.70]{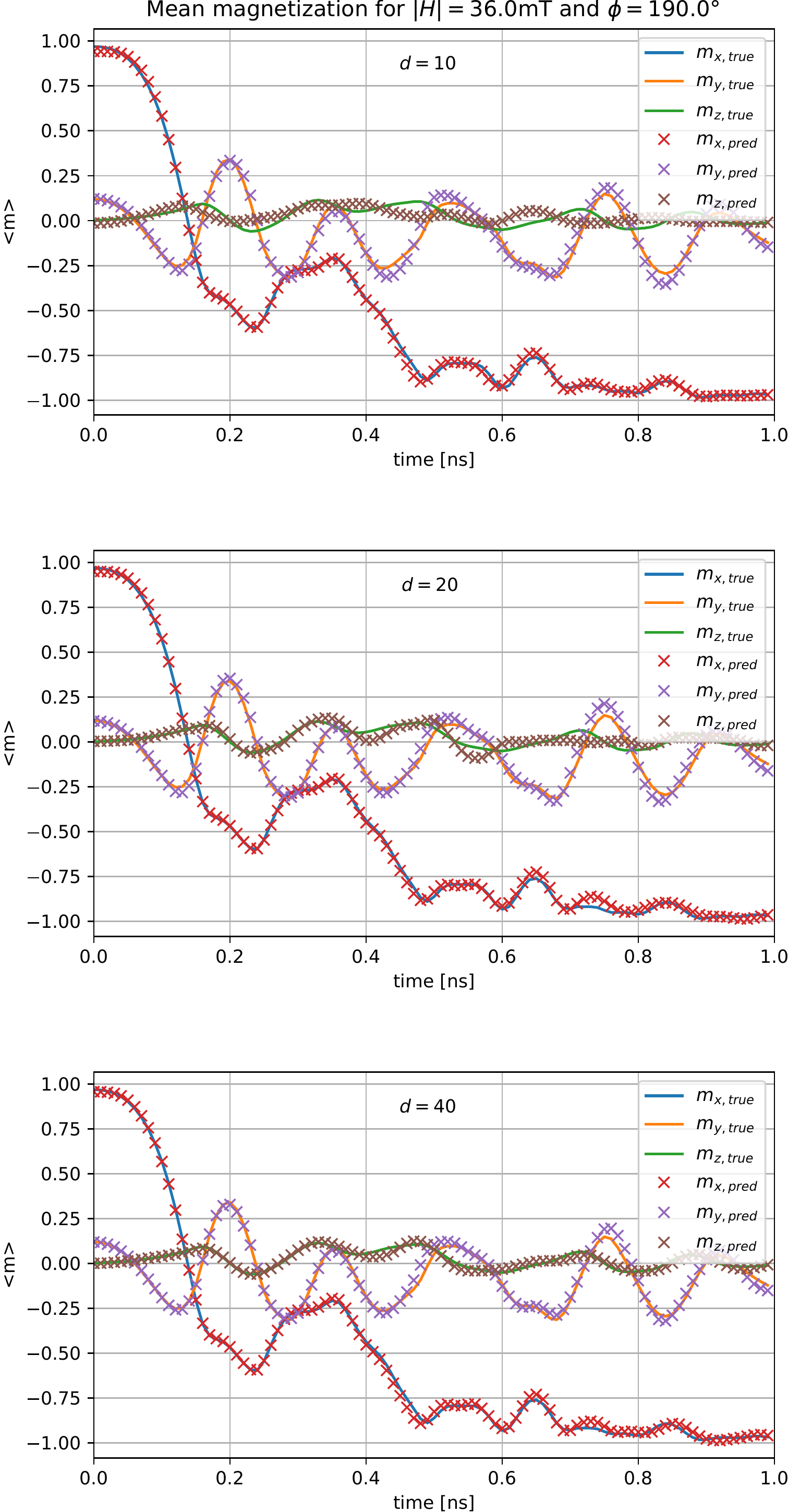}    
\caption{Predictions versus computed results for mean magnetization in the field $2$ case for varying number of kernel principal components $d = 10,20$ and $40$. The parameters were chosen as follows: kernel parameter $\gamma = 1/\luk{}{(3N)}$ and time-stepping $\nu=5$. A rank of $r=40$ was used. }\label{fig:field2_varying_d}
\end{figure}

\begin{figure}[hbtp]
\centering 
\includegraphics[scale=0.9]{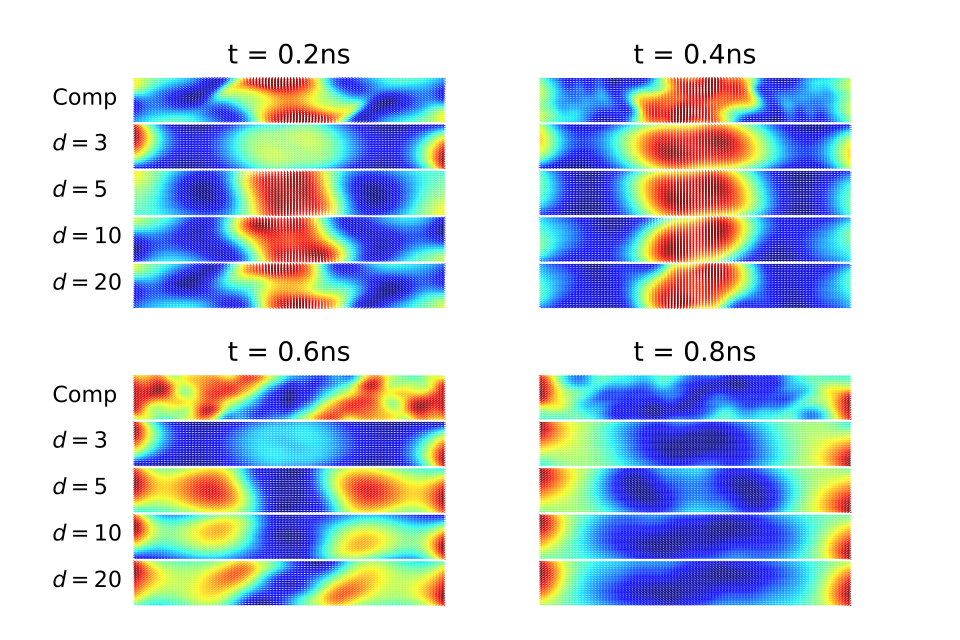}    
\caption{Snap shots of computed (Comp) and predicted magnetization states in the field $1$ case for varying number of kernel principal components $d = 3,5,10$ and $20$. The parameters were chosen as follows: kernel parameter $\gamma = 1/\luk{}{(3N)}$ and time-stepping $\nu=3$. A rank of $r=30$ was used.}\label{fig:field1_meanmag_d}
\end{figure}

\begin{figure}[hbtp]
\centering 
\includegraphics[scale=0.9]{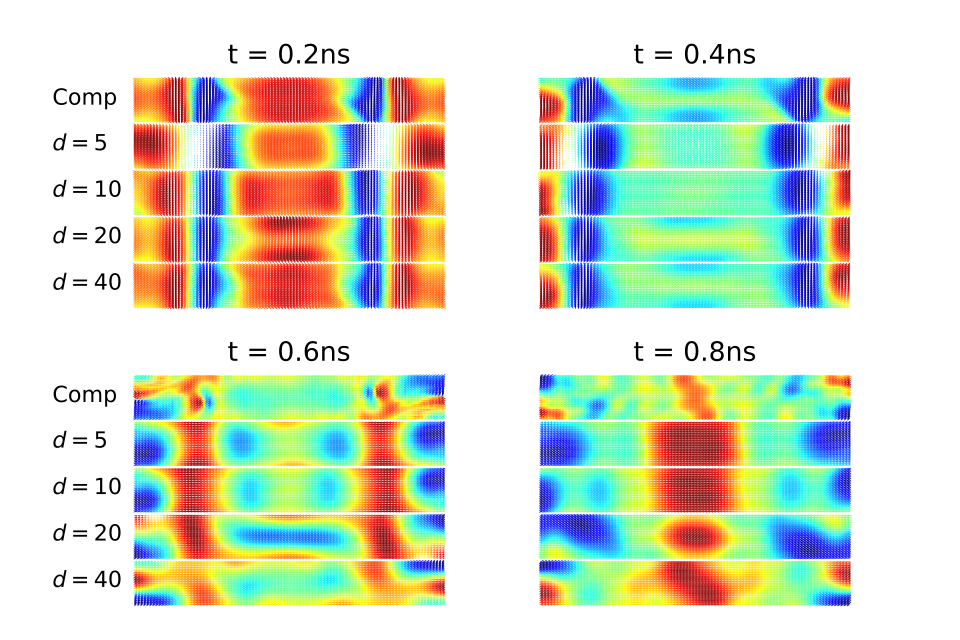}    
\caption{Snap shots of computed (Comp) and predicted magnetization states in the field $2$ case for varying number of kernel principal components $d = 5,10,20$ and $40$. The parameters were chosen as follows: kernel parameter $\gamma = 1/\luk{}{(3N)}$ and time-stepping $\nu=5$. A rank of $r=40$ was used.}\label{fig:field2_meanmag_d}
\end{figure}

Next we show the prediction in dependence of the number of kernel principal components $d$. Fig.~\ref{fig:field1_varying_d} compares the predictions of the mean magnetization dynamics with the computer simulations for $d=5,10$ and $20$ in the range of field $1$, and Fig.~\ref{fig:field2_varying_d} for $d=10,20$ and $40$ in the range of field $2$. We note that predictions of the trajectories take only a few seconds of computation time.
Fig.~\ref{fig:field1_meanmag_d} and Fig.~\ref{fig:field2_meanmag_d} illustrate snapshots of the predicted magnetization states in the two ranges depending on $d$.  

\begin{figure}[hbtp]
\centering 
\includegraphics[scale=.7]{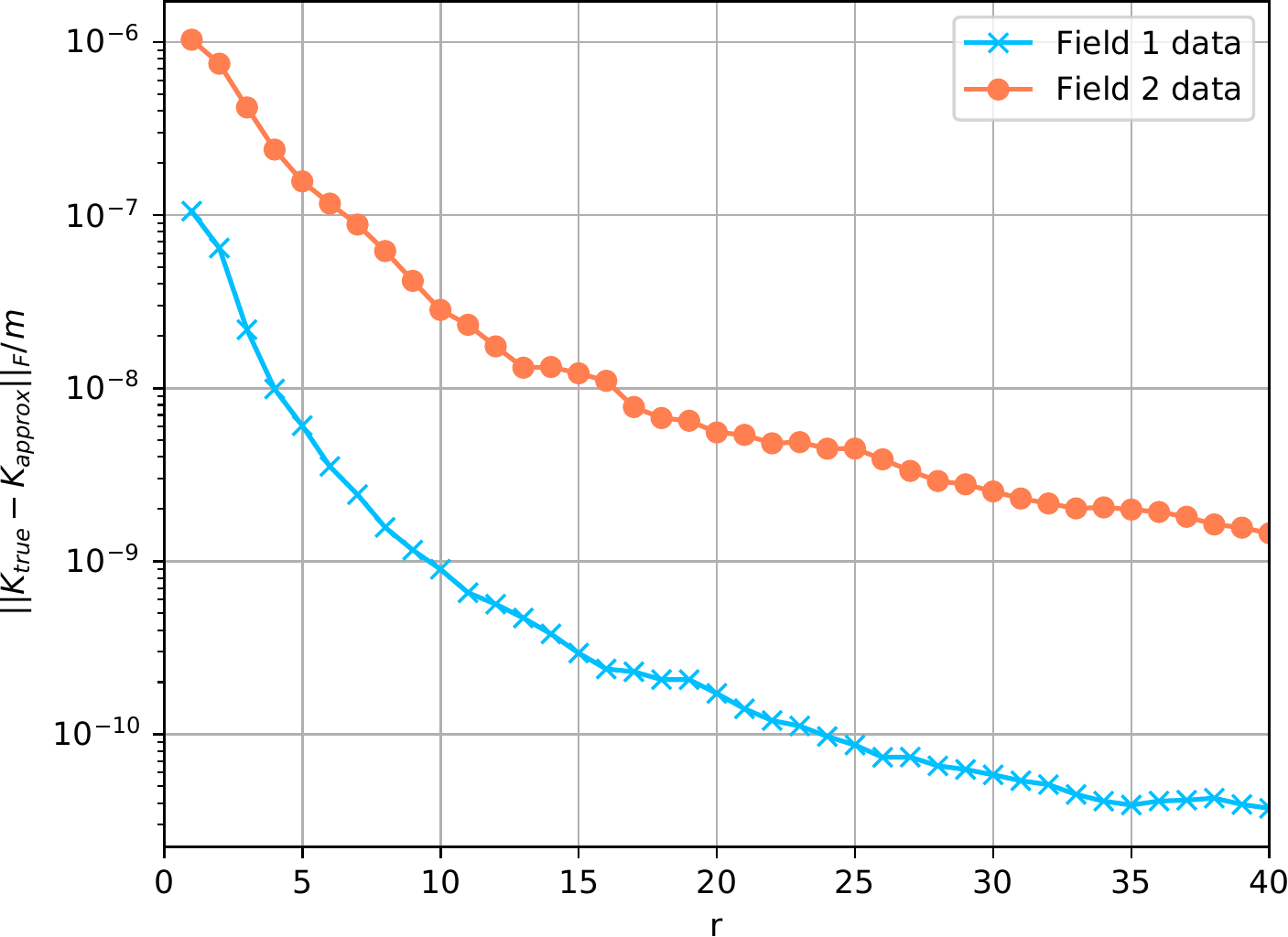}    
\caption{Low-rank kernel matrix approximation for increasing rank $r$ for the data sets corresponding to field 1 and 2, respectively.}\label{fig:kernel_decay}
\end{figure}

\vspace*{0.5cm}
\textbf{Low-rank variant.}\\
We validate the performance of the low-rank version of our proposed procedure. The previous validation indicates a choice of $\gamma = 1/\luk{}{(3N)}$ and e.g. $d=20$ and $\nu=3$ as sufficient in the field $1$ case, respectively $d=40$ and $\nu=5$ in the field 2 case. Fig.~\ref{fig:kernel_decay} shows the Frobenius error norm of the low-rank approximation of the kernel matrix for increasing rank $r$.   
%

\begin{figure}[hbtp]
\centering 
\includegraphics[scale=0.70]{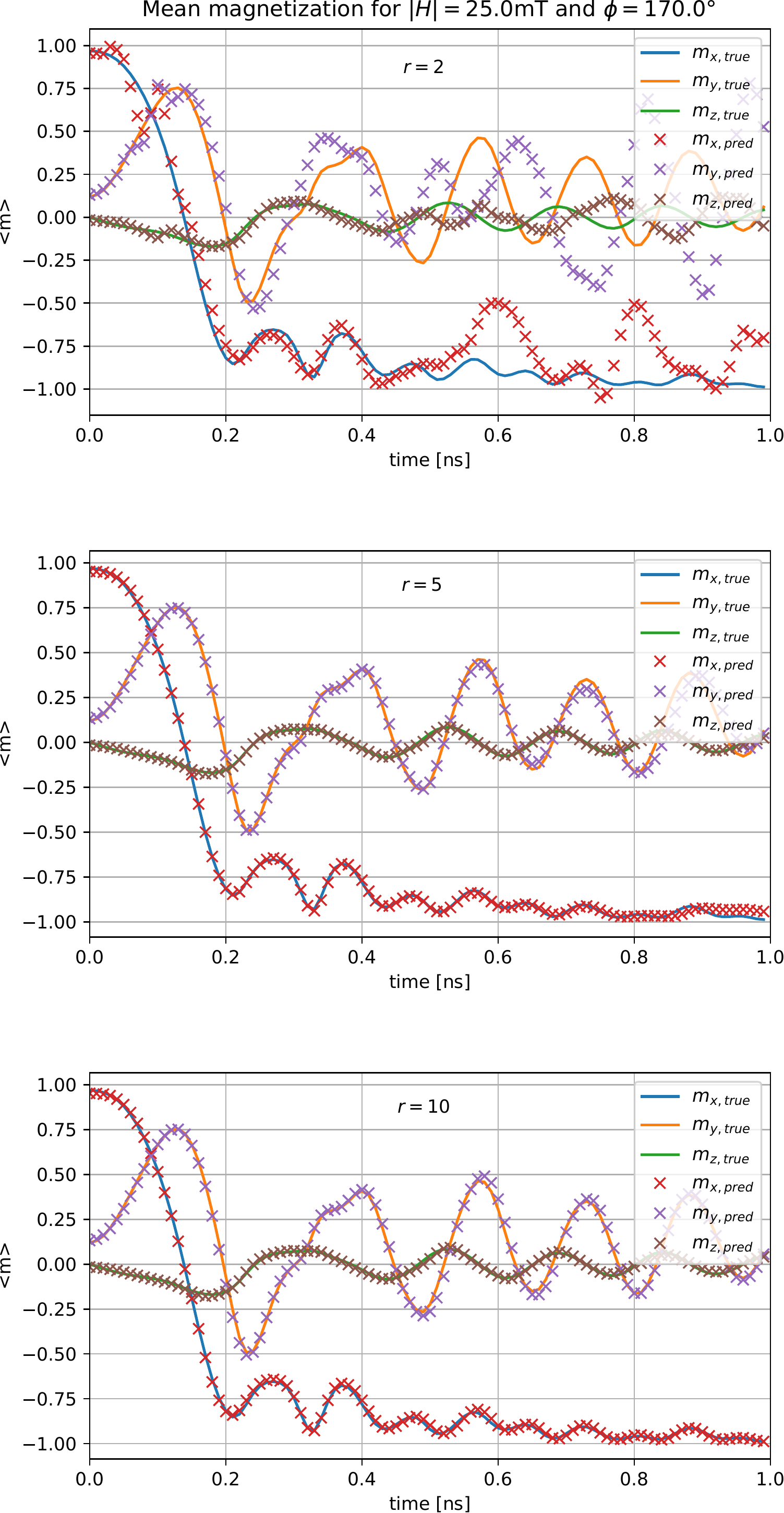}    
\caption{Predictions versus computed results for mean magnetization in the field $1$ case for varying rank $r = 2,5$ and $10$. The parameters were chosen as follows: kernel parameter $\gamma = 1/\luk{}{(3N)}$ and time-stepping $\nu=3$. A number of $d=20$ components was used.}\label{fig:field1_varying_r}
\end{figure}

\begin{figure}[hbtp]
\centering 
\includegraphics[scale=0.70]{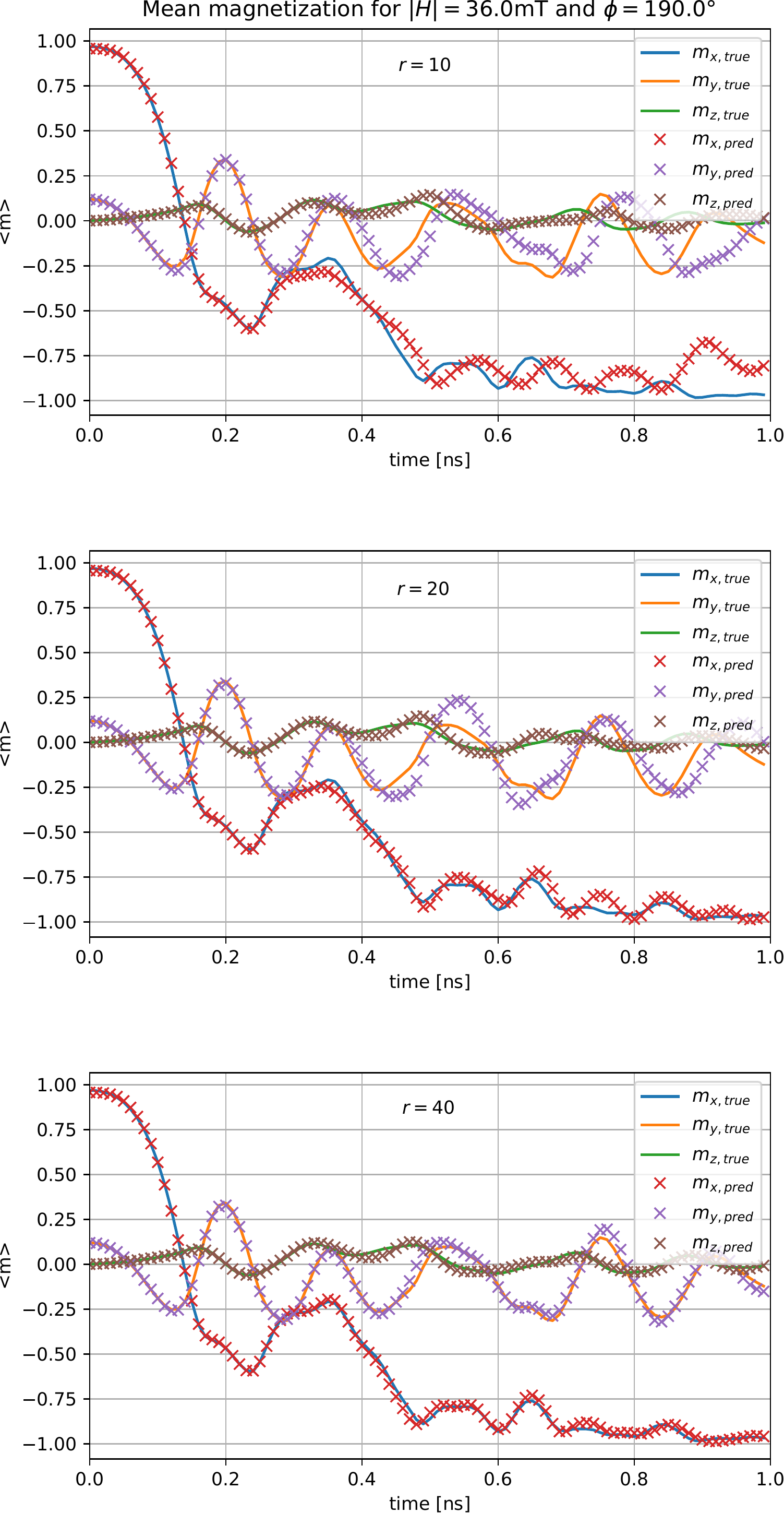}    
\caption{Predictions versus computed results for mean magnetization in the field $2$ case for varying rank $r = 10,20$ and $40$. The parameters were chosen as follows: kernel parameter $\gamma = 1/\luk{}{(3N)}$ and time-stepping $\nu=5$. A number of $d=40$ components was used. }\label{fig:field2_varying_r}
\end{figure}

\begin{figure}[hbtp]
\centering 
\includegraphics[scale=0.9]{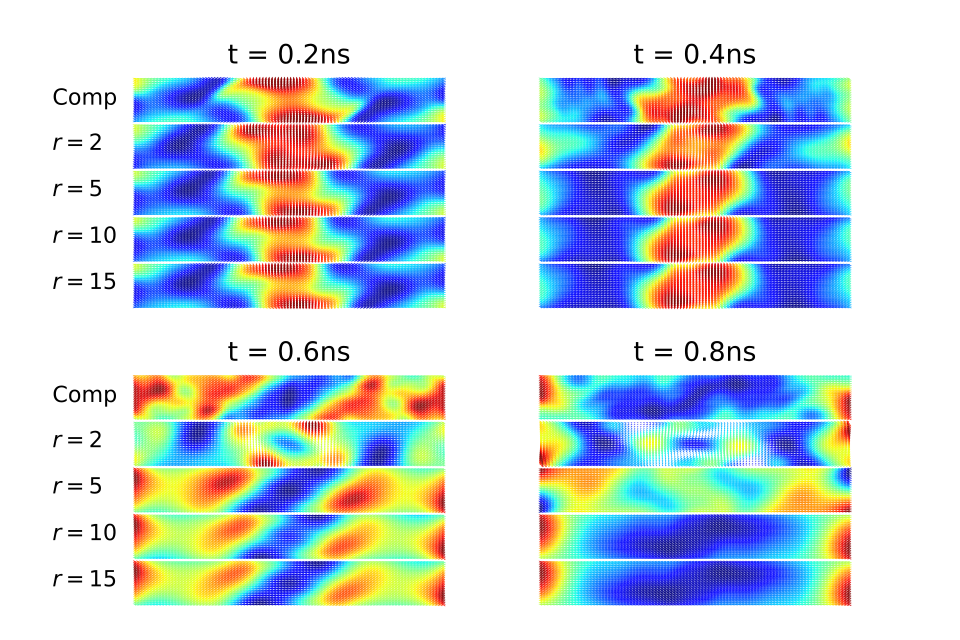}    
\caption{Snap shots of computed (Comp) and predicted magnetization states in the field $1$ case for varying rank $r = 2,5,10$ and $20$. The parameters were chosen as follows: kernel parameter $\gamma = 1/\luk{}{(3N)}$ and time-stepping $\nu=3$. A number of $d=20$ components was used.}\label{fig:field1_meanmag_r}
\end{figure}

\begin{figure}[hbtp]
\centering 
\includegraphics[scale=0.9]{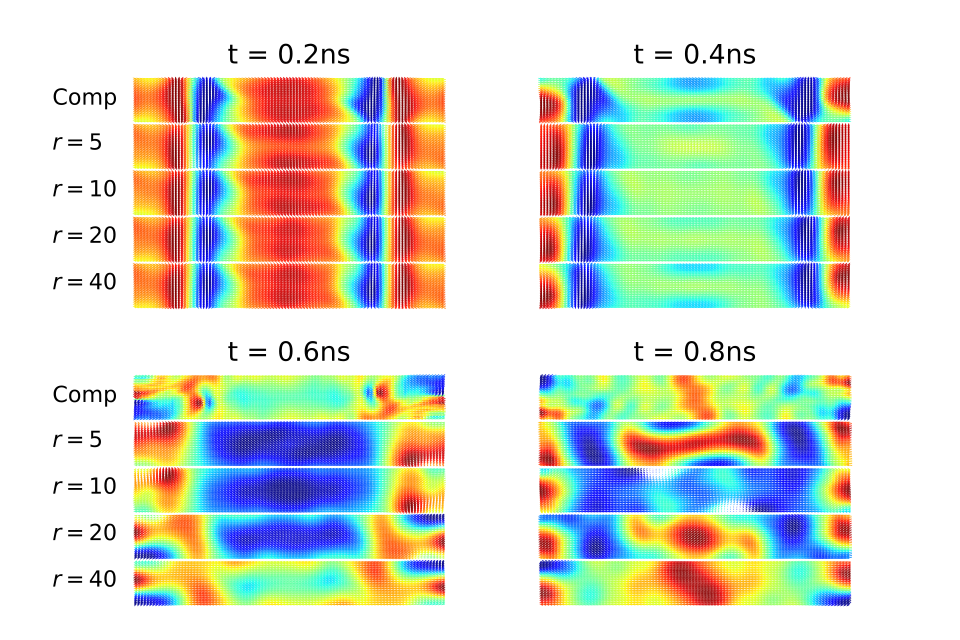}    
\caption{Snap shots of computed (Comp) and predicted magnetization states in the field $2$ case for varying rank $r = 5,10,20$ and $40$. The parameters were chosen as follows: kernel parameter $\gamma = 1/\luk{}{(3N)}$ and time-stepping $\nu=5$. A rank of $r=40$ was used.}\label{fig:field2_meanmag_r}
\end{figure}

Next we show mean magnetization plots and magnetization snap shots for increasing rank $r$. Fig.~\ref{fig:field1_varying_r} and Fig.~\ref{fig:field2_varying_r} show the mean magnetization for increasing $r$ in the field $1$ resp. the field $2$ case. Fig.~\ref{fig:field1_meanmag_r} and Fig.~\ref{fig:field2_meanmag_r} show associated magnetization snap shots.


The test cases on the NIST standard problem show the expected improvements in the predictions of mean magnetization curves and magnetization states for increasing number of kernel principal components, time stepping number as well as rank. However, for the less smooth manifold in the field 2 range \cite{exl2020learning} a clearly larger number of kernel principal components and rank is needed. 

\vspace*{0.5cm}
\luk{}{
\textbf{Computational costs.}\\
We compare the low-rank algorithm vs. the dense algorithm (no low-rank approximation), both in terms of training and prediction times as required for the magnetization plots in the field $2$ case 
as well as in terms of accuracy as given by the mean squared error at the final time point at $1$ns relative to the ground truth. Only the final pre-image is computed. As before, we use $d=40$ and $n=300$ and vary the rank $r$ and the time-stepping number $\nu$. We used a Intel(R) Core(TM) i7-4770K CPU \@ 3.50GHz.  Tab.~\ref{tab:cputimes} shows the respective results, where one can recognize a clear advantage in effort of the low-rank variant for the training and the prediction at comparable accuracy.}    
\begin{table}
\tabcolsep 6pt 
\caption{\luk{}{Cpu times in seconds for training and prediction of the whole dynamics of $1$ns in the field $2$ case for the low-rank variant for varying rank $r$ compared to the dense algorithm. The last column shows the mean squared error (mse) of the magnetic states at the final time point relative to the ground truth (only the final pre-image is computed). The number of field values is $n = 300$, and the number of components is $d=40$. We give data for time-stepping number $\nu = 3$ and $\nu=5$. }}\label{tab:cputimes}
\begin{center}
\begin{tabular}{c c c c c}
  & Algorithm &  training time &  prediction time &  mse \\
\hline\hline
 $\nu = 3$ & \textrm{dense} &              1184.56 &   4.64 &  0.031 \\ \hline
& \textrm{low-rank} ($r=10$) &  18.98 &   0.37 &  0.182 \\ \hline
& \textrm{low-rank} ($r=20$) &  39.21 &    0.97 &  0.092 \\ \hline
& \textrm{low-rank} ($r=30$) &  68.98 &    1.73 &  0.061 \\ \hline
& \textrm{low-rank} ($r=40$) &  110.45 &   2.64 &  0.041 \\ \hline\hline
$\nu = 5$ & \textrm{dense} &              1180.13 &   5.38 &  0.024 \\ \hline
& \textrm{low-rank} ($r=10$) &  18.95 &   0.46 &  0.161 \\ \hline
& \textrm{low-rank} ($r=20$) &  40.16 &     1.13 &  0.069 \\ \hline
& \textrm{low-rank} ($r=30$) &  71.41 &   2.02 &  0.037 \\ \hline
& \textrm{low-rank} ($r=40$) &  115.50 &   2.91 & 0.031 \\ \hline\hline
\end{tabular}
\end{center}
\end{table}

\section*{Conclusion}
We presented a low-rank version of kernel principal component analysis (low-rank kPCA) which utilizes a Nystroem approximation to the kernel matrix. 
The low-rank kPCA is capable of managing larger sets of training data. The key computational tasks in the low-rank kPCA, such as eigenvalue problems, projection onto kernel principal axes and the pre-image computation, are effectively treated by exploiting the low-rank structure of the Gram matrix. The low-rank kPCA was implemented as an extension in the scikit learn Python software \cite{scikitlearn}. 
We give a stand-alone validation example of the low-rank kPCA in the fashion of the scikit learn documentation. \luk{}{Training and prediction in the low-rank variant is shown to be significantly more effective than in the dense case.}
Following \cite{exl2020learning} we then apply the new method to establish a machine learning model to predict the micromagnetic dynamics described by the Landau-Lifschitz-Gilbert equation, the fundamental partial differential equation in mircomagnetics. Magnetization states from
simulated micromagnetic dynamics associated with different external fields are used as training data to learn a
dimension-reduced representation in feature space and a time-stepping map between the reduced spaces. 
The time-stepping prediction is based on learning maps between truncated representations of sample magnetization trajectories obtained by nonlinear model reduction via low-rank kPCA. Compared to the original proposed scheme in \cite{exl2020learning} the novel learning approach works entirely with reduced dimensional representations and the pre-image is only taken after the final time-step. The time stepping maps are established by a low-rank version of kernel ridge regression (low-rank kRR). Enhanced stability is observed when introducing multi-steps in the training process similar to \cite{kovacs2019learning}. We systematize this approach by incorporating a time-stepping number as hyper-parameter which we optimally determine via cross-validation, together with the number of kernel principal components. The test cases on the NIST standard problem show the expected improvements in the predictions of mean magnetization curves and magnetization states for increasing number of kernel principal components, time stepping number as well as rank. 
In principle, the proposed procedure allows to determine an "effective rank" during the low-rank approximation of the kernel matrix via the information obtained from the singular values. However, the selection of the basis vectors \luk{is not clear beforehand and should} could be systematized by procedures such as matching pursuit, possible future work but not yet treated in the present paper. \luk{}{A further limitation of the approach is the choice of the kernel hyperparameters which was accomplished by a very expensive cross validation, a procedure, which needs to be performed on only subsets of training data in case of very large problems. In addition the supervised approach is limited by the necessity of a large amount of conventional simulation results for the training data.} Future work shall also include application to other parameter-dependent differential systems such as nonlinear Schr\"odinger dynamics.

\section*{Acknowledgments}
We acknowledge financial support by the Austrian Science Foundation (FWF) via the projects "ROAM" under
grant No. P31140-N32 and the SFB "Complexity in PDEs" under grant No.
F65. We acknowledge the support from the Christian Doppler Laboratory Advanced Magnetic Sensing and
Materials (financed by the Austrian Federal Ministry of Economy, Family and Youth, the National
Foundation for Research, Technology and Development). The authors acknowledge the Wiener Wissenschafts und Technologie Fonds (WWTF) project No. MA16- 066 (“SEQUEX”) and the University of Vienna research platform MMM Mathematics - Magnetism - Materials. The computations were partly achieved by
using the Vienna Scientific Cluster (VSC) via the funded project No. 71140.
\bibliographystyle{abbrv}
\bibliography{bibref}

\end{document}